\documentclass[10pt,a4paper]{article}

\usepackage{amssymb}
\usepackage{amsmath}
\usepackage[all]{xy}
\usepackage{mathrsfs}
\usepackage{pstricks}
\usepackage{pst-3d}
\usepackage{pst-node}
\usepackage{pst-fill}
\usepackage{pst-coil}
\usepackage{graphicx}
\usepackage{lscape}

\newcommand{\Mathbf}[1]{\boldsymbol{#1}}
\newcommand{\qquadand}{\qquad\text{and}\qquad}

\newcommand{\df}{\frac{}{}}                              % Dummy fraction

\newcommand{\w}{\wedge}                                  % Wedge product
\newcommand{\wt}[1]{\widetilde{#1}}                       % Metric dual
\newcommand{\wh}[1]{\widehat{#1}}                         % Wide hat
\newcommand{\Lie}{\mathcal{L}}                           % Lie derivative
\newcommand{\diff}[2]{\frac{d{#1}}{d{#2}}}                  % Differentiation
\newcommand{\pdiff}[2]{\frac{\partial{#1}}{\partial{#2}}}   % Partial differentiation
\newcommand{\tensor}{\otimes}

\newcommand{\man}[1]{{\cal #1}}

\newcommand{\real}{\mathbb{R}}

\newcommand{\M}{\man{M}}
\newcommand{\ep}[1]{\epsilon_{#1}}
\newcommand{\g}{{g}}
\newcommand{\ginv}{\g^{-1}}
\newcommand{\cc}{c_{0}}
\newcommand{\Me}[1]{{\mathbf e}^{#1}}
\newcommand{\Mb}[1]{{\mathbf b}^{#1}}
\newcommand{\Md}[1]{{\mathbf d}^{#1}}
\newcommand{\Mh}[1]{{\mathbf h}^{#1}}
\newcommand{\Mm}[1]{{\mathbf m}^{#1}}
\newcommand{\Mp}[1]{{\mathbf p}^{#1}}
\newcommand{\ME}[1]{{\mathbf E}^{#1}}
\newcommand{\MB}[1]{{\mathbf B}^{#1}}
\newcommand{\MD}[1]{{\mathbf D}^{#1}}
\newcommand{\MH}[1]{{\mathbf H}^{#1}}
\newcommand{\MM}[1]{{\mathbf M}^{#1}}
\newcommand{\MP}[1]{{\mathbf P}^{#1}}
\newcommand{\V}{\mathbb{V}^U}
\newcommand{\Vt}{\wt{\mathbb{V}}^U}
\newcommand{\dotMb}[1]{\dot{\Mb{}}\hspace{-0.1cm}\phantom{i}^{#1} }

\newcommand{\dotMp}[1]{\dot{\Mp{}}\hspace{-0.1cm}\phantom{i}^{#1} }
\newcommand{\dotMB}[1]{\dot{\MB{}}\hspace{-0.1cm}\phantom{i}^{#1} }
\newcommand{\dotMD}[1]{\dot{\MD{}}\hspace{-0.1cm}\phantom{i}^{#1} }

\newcommand{\dotMM}[1]{\dot{\MM{}}\hspace{-0.1cm}\phantom{i}^{#1} }
\newcommand{\dotMP}[1]{\dot{\MP{}}\hspace{-0.1cm}\phantom{i}^{#1} }
\newcommand{\J}[1]{\man{J}^{U}_{#1}}

\newcommand{\s}[1]{\Mathbf{s}^{#1}}
\newcommand{\tauK}[1]{\tau_{K}^{#1}}
\newcommand{\JUK}[1]{J_{K}^{U#1}}
\newcommand{\rhoUK}[1]{\rho_{K}^{U#1}}
\newcommand{\PD}[1]{\partial_{#1}}
\newcommand{\sd}{\underline{d}}               % Spatial exterior derivative
\newcommand{\TA}[1]{\langle \; #1 \; \rangle} % Time average
\newcommand{\JU}[1]{J^{U}_{#1}}
\newcommand{\RU}[1]{\rho^{U}_{#1}}
\newcommand{\dotRU}[1]{\dot{\rho}^{\,U}_{#1}}
\newcommand{\hatRU}[1]{\wh{\rho}^{\,U}_{#1}}

 \newcommand{\BY}[1]{\textsc{#1},}                                 % Authors
 \newcommand{\atque}{{\normalfont and\;}}                       % Defining `and` for author macro
 \newcommand{\TITLE}[1]{\textit{#1},}                              % Title
 \newcommand{\IN}[4]{\textit{#1}, \textbf{#2}  (#3)  #4}     % Journal info: #1 - journal
\newcommand{\SEM}{stress-energy-momentum }
\newcommand{\EM}{electromagnetic }
\newcommand{\bfr}{{\mathbf r}}
\newcommand{\FF}{{\cal F}}
\newcommand{\pdxi}{\frac{\partial}{\partial x^i}}
\newcommand{\VV}{{\cal V}_t}
\newcommand{\TT}{{T^{tot}}}
\newcommand{\TENSOR}{{T}}
\newcommand{\Evec}{{\mathbf v}}
\newcommand{\Ewec}{{\mathbf w}}
\newcommand{\Mee}{{\underline{{\mathbf e}}^U}}
\newcommand{\Mbb}{{\underline{{\mathbf b}}^U}}

\newcommand{\vecJcal}{{\underline{{\mathbf {\cal J}}}^{U}}}
\newcommand{\Sym}{\text{Sym}}
\newcommand{\UUU}{{U,\M}}
\newcommand{\SIG}{{\Sigma_t}}
\newcommand{\inn}{{\it in}}
\newcommand{\outt}{{\it out}}
\newcommand{\sigOneTwo}{\Sigma_{12}}%{\sig12}{\Sigma_{12}}

\newcommand{\pmKill}[1]{\zeta_{#1}}
\newcommand{\eqq}{\;\;=\;\;}
\newcommand{\ESM}{appendix}

\begin{document}
%%%%%%%%%%%%%%%%%%%%%%%%%%%%%%%%%%%%%%%%%%%%%%%%%%%%%%%%%%%%%%%%%%%%%%%%%%%%%%%%%%%%%%%%%
\title{\textbf{The Electrodynamics of Inhomogeneous Rotating Media and the Abraham and Minkowski Tensors I: General Theory} }
\date{}
\author{Shin-itiro Goto, Robin W. Tucker and Timothy J. Walton \\
Department of Physics, Lancaster University, Lancaster and \\
The Cockcroft Institute, Keckwick Lane, Daresbury, UK.}

%%%%%%%%%%%%%%%%%%%%%%%%%%%%%%%%%%%%%%%%%%%%%%%%%%%%%%%%%%%%%%%%%%%%%%%%%%%%%%%%%%%%%%%%%
\maketitle

\begin{abstract}
This is paper I of a series of two papers, offering a
self-contained analysis of the role of \EM \SEM tensors in the
classical description of continuous polarizable perfectly
insulating media. While acknowledging the primary role played by
the {\it total} \SEM tensor on spacetime we argue that it is
meaningful and useful in the context of covariant constitutive
theory to assign preferred status to particular parts of this
total tensor, when defined with respect to a particular splitting.
The relevance of tensors, associated with the \EM fields that
appear in Maxwell's equations for polarizable media, to the forces
and torques that they induce has been a matter of some debate
since Minkowski, Einstein \& Laub and Abraham considered these
issues over a century ago. The notion of a force density that
arises from the divergence of these tensors is strictly defined
relative to some inertial property of the medium. Consistency with
the laws of Newtonian continuum mechanics demands that the {\it
total} force density on any element of a medium be proportional to
the local linear acceleration field of that element in an inertial
frame and must also arise as  part of the divergence of the {\it
total} \SEM tensor. The fact that, unlike the tensor proposed by
Minkowski, the divergence of the Abraham tensor depends {\it
explicitly} on the local acceleration field of the medium as well
as the electromagnetic field, sets it apart from many other terms
in the total \SEM tensor for a medium.

In this paper we explore how \EM forces or torques on moving media
can be defined covariantly in terms  of  a particular 3-form on
those spacetimes that exhibit particular Killing symmetries. It is
shown how the drive-forms associated with translational Killing
vector fields lead to explicit expressions for the \EM force
densities in stationary media subject to the Minkowski
constitutive relations and these are compared with other models
involving polarizable media in \EM fields that have been
considered in the recent literature.
\end{abstract}

%%%%%%%%%%%%%%%%%%%%%%%%%%%%%%%%%%%%%%%%%%%%%%%%%%%%%%%%%%%%%%%%%%%%%%%%%%%%%%%%%%%%%%%%%
\section{Introduction}
%%%%%%%%%%%%%%%%%%%%%%%%%%%%%%%%%%%%%%%%%%%%%%%%%%%%%%%%%%%%%%%%%%%%%%%%%%%%%%%%%%%%%%%%%
The interaction of matter with the electromagnetic field has
played a dominant role in the development of our understanding of
Nature. In classical Newtonian continuum mechanics one is
concerned with dynamic processes involving the interaction of
(bounded) classical continua with external forces and torques in
space (see e.g. \cite{KOV}).
%(see e.g. Kovetz 2000).
In relativistic continuum mechanics the theory is generalized to
incorporate the concept of the spacetime manifold and formulated
in terms of tensors on this manifold (see e.g. \cite{Maugin}).
%(see e.g. Maugin 1978).
In both formulations the histories of configuration variables must
be compatible with balance laws associated with possible
translational and rotational symmetries of a metric structure  and
an associated  balance law of `energy'. If electromagnetic fields
are involved these laws are supplemented with the macroscopic
Maxwell equations in media. As a result of cohesive forces
originating at the molecular level the interaction of a material
with these (and gravitational influences) results in locally
induced strains. These strains determine material stresses that
are encoded into various `stress-tensors'. These in turn describe
the local distribution of force and torque densities in the
medium. The mathematical structure of such tensors characterizes
the response of different materials to external influences. {\it
Constitutive relations} are relations that are used together with
field equations, equations of motion and boundary conditions to
fix the dynamic evolution of the independent dynamical variables.
If the configuration of the system (open or closed) involves
thermodynamic variables the classical laws of thermodynamics  can
be used to constrain these relations  for real media. In the
Newtonian formulation one uses the Maxwell and Cauchy stress
tensors \cite{Landau},
%(Landau \textit{et al.} 1984),
together with certain equations of state. In the relativistic
formulation a primary role is played by the total
stress-energy-momentum tensor for the system. The determination of
these tensors, together with appropriate constitutive relations
for different types of polarizable uncharged media often requires
input from experiment. The \EM constitutive properties and the
appropriate \EM \SEM for light in moving media has been a subject
of debate (and possible confusion) for over a century.
Difficulties arise in properly accounting for the local nature of
the wave-matter interaction in terms of experimentally measurable
effects. Even for static \EM fields there are (often non-linear)
subtle interactions that induce changes in shape or volume in
deformable media dependent on its thermodynamic state.

There is a considerable body of opinion suggesting that in some
sense the choice between different \SEM tensors describing
interactions of a material medium with an electromagnetic field is
a matter of convenience and that different choices simply provide
alternative descriptions of the same overall interaction\footnote{
The genesis  of this idea of classical duality may have its
origins in the analogy with wave-particle duality in quantum
mechanics.} \cite{Maugin1,GINZBERG,ISRAEL,Pfeifer}.
%(Ginzberg 1973; Israel 1977; Pfeifer \textit{et al.} 2007; Maugin 1980).
Since there is no preferred tensor partition of the {\it total}
\SEM tensor into sub-tensors describing the behaviour of
interacting {\it sub-systems} there can be no unique definition of
a \SEM tensor describing an interacting subsystem. Thus for \EM
fields interacting with an electrically  neutral bounded continuum
it is always possible to {\it redefine} integrated electromagnetic
forces and torques on the medium associated with sub-tensors
according to taste, particularly if the medium is composed of
piecewise inhomogeneous material subsystems or the fields are time
dependent. However in order to model the {\it total} interacting
system the choice of a total \SEM tensor is necessary. If one has
decided how to model a neutral but polarizable medium in the {\it
absence} of externally applied \EM fields then different choices
of a \SEM tensor for \EM fields in the medium to describe the
additional interactions with such fields will inevitable lead in
general to different predictions for the interacting system.

Although the laws of classical electromagnetism in the vacuum are
firmly established in a relativistic context there is no general
agreement on how best to accommodate the dynamics of deformable
media as a self consistent theory on spacetime, (see e.g.
\cite{REL-ELASTIC}).
%(see e.g. Frauendiener 2007).
This makes any rigorous formulation of
relativistic continuum mechanics of inhomogeneous dispersive
polarizable media interacting with \EM and gravitational fields
difficult even if one contemplates using it for systems in
non-relativistic motion in some frame of reference.

Despite these shortcomings there has recently been a resurgence of
interest in the so-called Abraham-Minkowski controversy and its
relevance to the use of either the Abraham or the Minkowski form
of \SEM tensor in interpreting experiments involving \EM fields in
media \cite{Hinds,ARCHIVE,HO,HOHO}.
%(Hinds \& Barnett 2009).
Part of this difficulty is no doubt due to the complex nature of
material responses to forces in general. From our perspective such
experiments are seeking constitutive relations involving {\it
particular} material systems and the macroscopic Maxwell fields in
media. As such it should come as no surprise that different
systems might yield different responses particularly if the
competing effects of electro- or magneto-striction mentioned above
are contributing differently in different experiments.

A number of historic experiments have sought to detect the
discriminating {\it Abraham force} \cite{WALKER2,WALKER1}.
%(Walker \textit{et al.} 1975; Walker \& Walker 1977).
This is strictly discriminatory only for homogeneous
non-dispersive {\it stationary} media, so calls into question
those experiments that involve media in motion. Indeed the
fundamental difference between the forces or torques induced by
the divergence of the Abraham and Minkowski \EM \SEM tensors is
that the former, unlike the latter, can depend {\it explicitly} on
the acceleration of the medium. From this observation it is our
contention that, for any specified \EM constitutive relation,
there remain experimental avenues offering new means to
discriminate between alternative proposals for the form of the \EM
\SEM tensor in media, despite the attendant inherent material
constitutive complexities. In particular we argue that, in
principle, the observed dependence of a time-averaged \EM
wave-induced torque on the angular speed of a {\it materially
isotropic but inhomogeneous} uniformly rotating electrically
neutral medium could be used to discriminate between the \EM wave
interactions described by the Abraham tensor from those described
by the symmetrized version of the tensor introduced by Minkowski,
and possibly those proposed by others.

It is therefore of interest to calculate how such a torque depends
on geometric properties of a  cylindrical insulator, its speed of
rotation and its {\it \EM constitutive properties}. Since the
medium will be assumed electrically polarizable and magnetizable
(but non-conducting) one is immediately confronted with a number
of subtleties associated with the form of this tensor and the {\it
material constitutive properties} of the medium. These issues have
a bearing on how one formulates the classical electromagnetic
force on a macroscopic body particularly one that is accelerating.
Since this has led to a number of related questions in the recent
literature \cite{BREVIK,Barnett,Mans},
%(Brevik 1970; Barnett \& Loudon 2006; Mansuripur 2008),
this article attempts to make explicit our perspective. After
reviewing approaches in the Newtonian framework the use of a fully
covariant special relativistic framework is advocated. The
essentials of this formulation are described below which should be
read in conjunction with the expository material on
electromagnetic theory in the language of differential forms
contained in the appendix. In the following paper, applications to
rotating media are discussed in this context.

%%%%%%%%%%%%%%%%%%%%%%%%%%%%%%%%%%%%%%%%%%%%%%%%%%%%%%%%%%%%%%%%%%%%%%%%%%%%%%%%%%%%%%%%
\section{Points of Departure}\label{DEPARTURE}
%%%%%%%%%%%%%%%%%%%%%%%%%%%%%%%%%%%%%%%%%%%%%%%%%%%%%%%%%%%%%%%%%%%%%%%%%%%%%%%%%%%%%%%%%
In order to motivate our method of analysis leading to a
computation of the \EM torque on an inhomogeneous rotating
uncharged insulator, it is useful to place our methodology in the
context of the recent literature in this subject. Some of this
literature is devoted to the derivation of expressions for the
classical force (and torque) induced by \EM fields on electrically
neutral polarizable continua based on an underlying discrete model
of \EM sources. A traditional non-relativistic approach is to take
as a point of departure the classical vacuum Maxwell equations for
the fields $\Me{U}(\bfr,t)$ and $\Mb{U}(\bfr,t)$ in some inertial
frame (here labelled $U$) and moving point sources\footnote{We
assume throughout that free magnetic charge is absent in Nature.}
together with the Newtonian equations of motion in $\real^{3}$ for
the sources. For a collection of $N$ point charges where the
$\alpha$-th point has mass $m_{\alpha}$ and Newtonian velocity
${\mathbf v}_{\alpha}(t)$ at time $t$, the motion of each particle
is given by a solution to the $N$ coupled ordinary differential
equations:
\begin{eqnarray}\label{NEWT}
    \diff{}{t}\left(\df m_{\alpha} {\mathbf v}_\alpha(t) \right) \eqq \wh{\FF}_{\alpha}(t) + \sum^{N}_{\beta\neq\alpha} \FF_{\alpha\beta}(t),\qquad\qquad \alpha,\,\beta=1\ldots N
\end{eqnarray}
where $\FF_{\alpha\beta}(t)$ is the inter-particle force and
$\wh{\FF}_{\alpha}(t)$ the resultant force on the $\alpha$-th
particle due to all other influences. The non-relativistic
interaction of the ambient Maxwell fields $\Me{U}$ and $\Mb{U}$
with any point particle with charge $q_{\alpha}$ located at
$\bfr=\bfr_{\alpha}(t)$  with Newtonian velocity $ {\mathbf
v}_\alpha(t) =\dot{\bfr}_{\alpha}(t) $ is derived from the
electromagnetic force $\FF_{\alpha}^{EM}(t)$ on that particle
with:
\begin{eqnarray}
\label{LF}
\FF_{\alpha}^{EM}(t) \eqq q_{\alpha} \left( \Me{U}(\bfr_{\alpha}(t),t) + {\mathbf v}_{\alpha}(t) \times \Mb{U}(\bfr_{\alpha}(t),t)\right)
\end{eqnarray}
in the Gibbs vector notation. The fields themselves are in turn
derived from the {\it vacuum microscopic } Maxwell equation with singular
sources. In the Gibbs {\it vector field notation} these are the
equations
\begin{eqnarray}
 \text{curl}\, \Mee &=& -\pdiff{\Mbb}{t} \\
 \text{div} \,\Mbb &=& 0 \\
{\mu_0}^{-1}\text{curl}\,\Mbb &=& \vecJcal +  \ep{0}\pdiff{\Mee}{t} \\
\ep{0}\,\text{div}\,\Mee &=& \wh{\rho}^{U}
\end{eqnarray}
respectively, where $\mu_{0}^{-1}=\cc^{2}\ep{0}$, $\cc$ denotes
the speed of light in vacuo, $\wh{\rho}^{U}= \sum_{\alpha} \,
q_{\alpha} \delta^{D}( \bfr - \bfr_{\alpha}(t) )$ and $\vecJcal =
\sum_{\alpha} q_{\alpha}{\mathbf v}_{\alpha}(t) \delta^{D}( \bfr -
\bfr_{\alpha}(t) )$ in terms of the singular Dirac distribution
$\delta^{D}$. In these equations only the fundamental \EM fields
$\Me{U}$ and $\Mb{U}$ play a role. Multi-pole magnetization and
polarization sources are defined in terms of the attributes of the
charged particles  and their motion as limits in a multi-pole
expansion. In non-relativistic models retardation effects due to
the Maxwell displacement current are often ignored. It is
straightforward to verify that the total {\it linear particle
momentum} in the system, $\sum_{\alpha}^{N} \,m_{\alpha} {\mathbf
v}_{\alpha}(t)$, is a constant of the motion provided
$\sum_{\alpha}^{N} \,\wh{\FF}_{\alpha}(t)=0$ and
$\FF_{\alpha\beta}(t)= -\FF_{\beta\alpha}(t)$. From these
fundamental assumptions one may approach a continuum description
from a number of distinct directions including a multi-pole
expansion of the \EM fields about some arbitrary point in space
followed by a spatial smoothing procedure for the singular
sources, in order to generate a balance law for smoothed out total
non-relativistic linear momentum for the total interacting system.
This yields an Euler continuum description from the fundamental
particle-field description. While the resulting overall continuum
balance equation may not be sensitive to the precise nature of the
smoothing functions the interpretation of individual forces in the
Eulerian balance relation may be \cite{GROOT,MURDOCH}.
%(de Groot \& Suttorp 1972; Murdoch \& Bedeaux 1994).

An alternative approach to a continuum description has been to
assume that the point particles are electrically neutral (atoms or
molecules) but are endowed with elementary electric dipole moments
and/or magnetic dipoles or elementary current loops. This requires
that (\ref{LF}) be changed to reflect this modification.
Subsequent smoothing would lead to a balance law involving force
terms different from the point charge model even in the static
limit. This should come as no surprise since the underlying
microscopic models are different.

In comparing the predictions of different continuum models one
must be aware that they may differ only in their effects at the
boundaries of spatially compact media. Terms that may be discarded
during an integration by parts in the development of the modelling
process may well contribute to boundary interactions depending on
the nature of the boundary conditions.

To illustrate some of these points consider the typical modelling
of a homogeneous dispersion-free electrically neutral stationary
medium \cite{SCHWINGER} %(Schwinger \textit{et al.} 1998)
based on a collection of point charges yielding, in some continuum
limit, a time dependent volume force density on $\real^{3}$ given
by
\begin{eqnarray}
\Mathbf{F}_{t} \eqq \frac{1}{2}\nabla\left( \Me{U} \!\cdot \Mp{U} + \Mb{U} \!\cdot \Mm{U} \right) +  \frac{1}{2}\pdiff{}{t}\left( \Mp{U}\times\Mb{U}\right) \eqq F_{1t}\Mathbf{i} + F_{2t}\Mathbf{j} + F_{3t}\Mathbf{k}\label{SWIG},
\end{eqnarray}
expressed in terms of time dependent \EM vector fields on
$\real^{3}$ in a global orthonormal Cartesian frame with basis
$\{\Mathbf{i,j,k}\}$ and Cartesian coordinates $\{x^{1},x^{2},x^{3}\}$.
The force vector on a finite volume $\VV$ of the medium is then
\begin{eqnarray*}
    \bar{\Mathbf{F}}_{t} &=& \int_{\VV}  \Mathbf{F}_t\; dx^{1}\, dx^{2}\, dx^{3}
\end{eqnarray*}
In (\ref{SWIG}), the $\Mp{U}$  and $\Mm{U}$ are spatially smoothed
time-dependent electric and magnetic dipole fields on $\real^{3}$
obtained by truncating a particle force multi-pole expansion.
Prior to smoothing they can be expressed in terms of the charge,
position and velocity of the electrically charged constituents of
the medium relative to some arbitrary point and the bulk  velocity
in the medium. They are superscripted to indicate that these
fields are referred to an inertial (laboratory) reference frame
$U$. In principle $\Mp{U}$  and $\Mm{U}$ are determined from the
continuum limit of the particle equations of motion. In practice
this is difficult so one resorts to constitutive relations and
assumes a form for the bulk motion of the continuum. With these
closure relations the vacuum Maxwell system absorbs the smoothed
polarization and magnetization sources into  the phenomenological
fields $\Mp{U}$, $\Mm{U}$, $\Md{U}$ and $\Mh{U}$.

In view of the covariant language used in the rest of this paper,
it is useful to write (\ref{SWIG}) in tensor form using the
Killing symmetry of the Euclidean structure defined by the metric
of 3-dimensional space. The reader should consult the \ESM for a
self-contained formulation of the Maxwell system in terms of the
frame dependent differential 1-forms $ \Me{U}, \Mb{U},  \Md{U},
\Mh{U}, \Mp{U}, \Mm{U}$ on space and their definition  in terms of
a unit time-like vector field $U$ and the Maxwell and excitation
2-forms $F$ and $G$ respectively on 4-dimensional spacetime
endowed with the Lorentzian metric tensor field $\g$\footnote{
    All electromagnetic tensors in this article have
    dimensions constructed from the SI dimensions $[M], [L], [T], [Q]$ where $[Q]$
    has the unit of the Coulomb in this system. We adopt $[\g]=[L^{2}],
    [G]=[j]=[Q],\,[F]=\frac{[Q]}{[\ep{0}]}$ where the permittivity of free space $\ep{0}$ has the dimensions $[ Q^{2} T^{2} M^{-1} L^{-3}]$.
    Note that the operators $d$ and $\nabla$ defined in the \ESM preserve the physical dimensions of tensor fields but with $[\g ]=[L^{2}]$, for $p$-forms $\alpha$ in $4$ dimensions,
    one has $[\star \alpha]=[\alpha] [L^{4-2p}]$.}. In a Cartesian coordinate system on $\real^{3}$, the covariant and contra-variant metric tensor fields are
\begin{eqnarray*}
    {\mathbf \g} \eqq \sum_{i=1}^{3}\,dx^{i} \tensor dx^{i}, \qquad\text{with}\qquad {\mathbf \ginv} \eqq \sum_{i=1}^{3}\,\pdxi \tensor \pdxi.
\end{eqnarray*}
and in these coordinates the associated translational (Killing)
vector fields are $\{\pdiff{}{x^{i}}\}$ (for all time $t$)
satisfying $ {\cal L}_{\pdiff{}{x^{i}}} {\mathbf \g} =0$. If $K$
is any of these Killing fields,
then\footnote{For any metric tensor field $\mathcal{G}$, we define the $\mathcal{G}$-dual of any 1-form $\alpha$ to be the vector field $\wt{\alpha}=\mathcal{G}^{-1}(\alpha,-)$. Similarly, for any vector $V$, we define the $\mathcal{G}$-dual 1-form $\wt{V}=\mathcal{G}(V,-)$.}
$\wt{K}=dx^{i}$, for some
Cartesian coordinate $x^{i},\; i=1,2,3$. In terms of
$\kappa=\#\wt{K}$, with $\#1=dx^{1} \w dx^{2} \w dx^{3}$, a
Euclidean covariant representation of (\ref{SWIG}) that meshes
with the conventions to be established below  is given in terms of
the equivalent force 3-form for direction $K$:
\begin{eqnarray}\label{MODEL}
    \FF_{K} &=&  \frac{1}{2}\left[ \sd\left( \Me{U}\!\cdot\Mp{U} + \Mb{U}\!\cdot\Mm{U} \right) + \# \Lie_{\PD{t}}\left( \Mp{U} \w \Mb{U}\right)  \right] \w \kappa,
\end{eqnarray}
where we define $\alpha\cdot\beta = \ginv\left(\alpha,\beta\right)
= {\mathbf \ginv}\left(\alpha,\beta\right)$ for any {\it{spatial}}
1-forms $\alpha,\beta$ and $\sd$ denotes the spatial exterior
derivative\footnote{See \ESM for the definition of the spatial
exterior derivative $\sd$ in terms of the exterior derivative $d$
on spacetime, and for definitions of other spatial operators.} on
$\real^{3}$. Then the Cartesian component of the {\it vector
force} in the Cartesian direction $i$ on a volume $\VV$ due to its
surroundings and ambient fields in the medium is given by
\begin{eqnarray*}
    \FF_{i}\left[\VV\right] &=& \int_{\VV} \FF_{\PD{i}}.
\end{eqnarray*}
Note that in the static field situation the force 3-form
(\ref{MODEL}) is expressed in terms of the gradient of the
interaction scalar $\frac{1}{2}(\Me{U}\!\cdot\Mp{U} +
\Mb{U}\!\cdot\Mm{U})$. Using
%\begin{eqnarray*}
%{\color{red}\sd\,\Me{U} = -\dotMB{U}, }
%\end{eqnarray*}
%(\ref{M1}),
the macroscopic Maxwell equations
\begin{eqnarray}
     \label{M1} \sd\,\Me{U} &=& -\dotMB{U} \\
     \label{M2} \sd\,\MB{U} &=& 0 \\
     \label{M3} \sd\,\Mh{U} &=& \J{} + \dotMD{U} \\
     \label{M4} \sd\,\MD{U} &=& \RU{}
\end{eqnarray}
where $\MD{U}=\#\Md{U}$, $\MB{U}=\#\Mb{U}$, $\dot{\alpha}=\cc\Lie_{\partial_{t}}\alpha$ for any $p$-form $\alpha$
and with the definitions
%(\ref{polarJrho}),
\begin{eqnarray}\label{polarJrho}
    \frac{\J{p}}{\cc} = -i_{U}j_{p} = \frac{1}{\cc}\left( \dotMP{U} - \sd\Mm{U}\right) \quad \text{and} \quad \RU{p} = -(i_{U}\star j_{p})\star\wt{U} = -\sd\MP{U},
\end{eqnarray}
one may write\footnote[1]{See {\ESM}
%appendix~\ref{SchwingerSect}
for the definition of the Hodge maps $\#$ and $\star$, and further
details of this calculation.}, with $\MP{U}=\#\mp{U}$ and
$\ME{U}=\#\Me{U}$:
\begin{eqnarray}
\FF_{K} = \tfrac{1}{2}\!\left[\RU{p}\Me{U}\!(K) + i_{\wt{\Mb{U}}}\J{p} \!\w \kappa + \Lie_{\wt{\Mb{U}}}\Mm{U}\! \w \kappa + \Lie_{K}\Mp{U} \!\w \ME{U} + \sd\!\left( \Me{U}(K)\MP{U} \right)\right]\label{MODEL1}.
\end{eqnarray}
Although the overall charge of the medium is taken to be zero the
first two terms in the equation for $\FF_{K}$ above constitute a
local Lorentz force density\footnote[2]{The factor $ \frac{1}{2}$
arises since the polarization is smoothly distributed in the
medium.} generated by the {\it induced polarization} charge
$\RU{p}$ and current $\J{p}$. The exact 3-form in the last term
will contribute to forces on the spatial boundary of the medium
unless they happen to be zero as a result of boundary conditions
satisfied by the fields there. The remaining terms describe local
forces depending on inhomogeneities  arising from the spatial rate
of change of polarization and magnetization in the system.

The above derivation of the structure of a local Newtonian force
density in a neutral macroscopic continuum in an \EM field starts
with a particular non-relativistic model in an inertial frame. We
stress that in the absence of dynamical information about the
polarization and magnetization the cogency of (\ref{MODEL1})
demands a knowledge of supplementary constitutive relations to
accommodate the response of the medium (via $\Mp{U}$ and $\Mm{U}$)
to the fundamental fields $\Me{U}$ and $\Mb{U}$. One could proceed
to express this force density as the sum of the spatial divergence
of a time dependent Maxwell second rank stress tensor on
$\real^{3}$ and a time derivative modulo boundary
terms\footnote[3]{There is no unique way to perform such a
split.}. Adding the Cauchy tensor describing the medium in the
absence of \EM forces to such a Maxwell stress tensor would give
the total Cauchy stress for the medium (in the presence of fields)
that enters into the local balance law for non-relativistic linear
momentum for the complete interacting system. The classical local
dynamics of this system follows as a solution to such a balance
law and requires implementation of interface conditions at media
boundaries or interfaces where material properties of the
continuum are discontinuous. The latter constraints benefit from a
distributional reformulation that has been  discussed more fully
elsewhere \cite{GCM8,RWT_JMP,RWT_YORK}.
%(Tucker \& Walton 2009; Tucker 2009, 2010).

An alternative point of departure for the modelling process is to
start with a fully covariant total symmetric \SEM
tensor $\TT$ for a medium interacting with classical fields on a general spacetime\footnote[5]{In the SI system the tensor $\TT$ has the physical dimension of a force, i.e. $\frac{M\,L}{T^{2}}$.}.
The fully covariant local classical equations of motion of the
medium are then postulated to be given by
\begin{eqnarray}\label{DIVT}
    \nabla\cdot\TT &=& 0
\end{eqnarray}
together with equations for the fields. Here the divergence operator $\nabla\cdot$ is defined with respect to the Levi-Civita connection $\nabla$  on spacetime. The fundamental property
of $\TT$ is that it acts as a source of Einstein gravitation. In
most classical considerations one demands $\TT(U,U)>0$ for all
future pointing timelike unit vector fields $U$, reflecting the
attractive nature of gravitation and $\TT(U,U)$ is then identified
with  local mass-energy density in the frame $U$. In a source-free
region of spacetime containing only \EM fields this condition is
maintained and identifies field energy.

In a spacetime with local isometries generated by a set of Killing
vector fields associated with the spacetime metric $\g$, $\TT$ can
be used to generate a set of closed 3-forms on spacetime. The
vanishing of the exterior derivative of each form in this set can
in turn give rise to a conservation law when integrated over a
regular 4-dimensional domain of spacetime provided the forms are
free of singularities there. Then a class of Killing vectors that
generate spatial translations  can be used to construct
conservation laws that reduce to the balance laws for the
components of Newtonian linear momentum in some non-relativistic
Newtonian limit.

Thus if $\TENSOR$ is {\it any} symmetric rank 2 symmetric tensor
on spacetime so $\TENSOR_{ab}=\TENSOR_{ba}$ where
\begin{eqnarray*}
    \TENSOR &=& \TENSOR_{ab}\, e^{a}\tensor e^{b}
\end{eqnarray*}
in any cobasis of 1-forms $\{e^{a}\}$ with a dual basis of vector fields
$\{X_{b}\}$ one defines the {\it drive 3-form} associated with
$\TENSOR$ and $K$:
\begin{eqnarray}\label{DRIVES}
    \tauK{} &=& -\pmKill{K}\,\TENSOR(K,X_{a})\star e^{a}
\end{eqnarray}
for any vector field $K$ on spacetime where $ \pmKill{K} =\pm 1 $
will be defined below to conform with the physical interpretation
of the different components of the drive-form. Given a frame
defined by the unit timelike (future pointing) observer
field\footnote{The frame is inertial if $\nabla U=0$.} $U$, one may decompose $\tauK{}$ into a {\it
spatial} 2-form $\JU{K}$ and {\it spatial} 3-form $\RU{K}$
relative to $U$ on spacetime:
\begin{eqnarray}\label{tau_split}
    \tauK{} &=& \JU{K} \w \wt{U} + \RU{K}
\end{eqnarray}
with $i_{U}\JU{K}=i_{U}\RU{K}=0$. When $K$ is a Killing vector
field ($\Lie_{K}\g=0$) it is a mathematical identity
%\cite{RWT,Benn}
(Benn \& Tucker 1988; Benn 1982)
that\footnote{More generally if $T$ is an
arbitrary $(2,0)$ tensor, $T=T_{ab}\,e^{a} \tensor e^{b}$, of no
particular symmetry, one has for any vector field $W$ the identity
$\frac{1}{2}( \Lie_{W}\g )(X_{a},X_{b})\,T^{\{ab\}} \star 1 =
-\pmKill{W}\,d\,\tau_{W} - (\nabla \cdot (\Sym\,T))(\wt{W})\star 1$
where $-\pmKill{W}\,\tau_{W}=(\Sym\,T)(W, X_{a} )\,\star e^{a}$ and
$\Sym\,T$ is the symmetric part of $T$ with components $T_{\{ab\}}
= \frac{1}{2}(T_{ab} + T_{ba})$.}
\begin{eqnarray}\label{THEOREM}
    d\tauK{} &=& -\pmKill{K}\, (\nabla\cdot T)(K) \star 1.
\end{eqnarray}
Hence, if $\nabla\cdot\TENSOR=0$ then
\begin{eqnarray}\label{DTAU}
    d\tauK{} &=& 0.
\end{eqnarray}
In terms of $\JU{K}$ the conservation equation (\ref{DTAU})
becomes
\begin{eqnarray*}
    d\,\JU{K} + \Lie_{U}\tauK{} &=& 0.
\end{eqnarray*}
If $K$ is a {\it spacelike translational} Killing vector field
with open integral curves then
\begin{eqnarray*}
    \JU{K} &\equiv& -i_{U}\tauK{}
\end{eqnarray*}
is a linear momentum current (stress) 2-form in the frame $U$
and
\begin{eqnarray*}
    \RU{K} &\equiv& -(i_{U}\star\tauK{})\star \wt{U}
\end{eqnarray*}
is the associated linear momentum density 3-form in the frame $U$.
If $K$ is a {\it spacelike rotational} Killing vector field
generating $SO(3)$ group isometries with closed integral curves,
then $\JU{K}$ is an angular-momentum current (torque stress)
2-form and $\RU{K}$ is the associated angular-momentum density
3-form in the frame $U$. If $K$ is a {\it timelike translational}
Killing vector field, then $\JU{K}$ is an energy current (power)
2-form and $\RU{K}$ is the associated energy density 3-form in the
inertial frame field  $U$. In the following, attention will be
directed mainly to particular translational and rotational
spacelike Killing vectors $K$ of flat spacetime and the
computation of integrals of $\JU{K}$  and $ \RU{K}$ for a
particular contribution to $\tauK{}$ associated with
electromagnetic fields in media undergoing various states of
motion observed in an inertial frame defined by $U$.

As noted above when $\TT$ describes a domain of vacuum spacetime,
free of matter but containing electromagnetic fields, one requires
the \EM  field energy in any local frame $U$ to be positive, i.e.
$\#\rho^U_U > 0$ in terms of $\rho^U_U$. Furthermore we shall
require that, for any spacelike Killing vector field $K$,
$\#\rho^U_K >0$. This ensures that when $K$ generates spatial
translations in Minkowski spacetime  and $U$ defines an inertial
frame then the time-averaged integral of $\rho^{U}_{K}$ over a
finite region of space can be identified with the time-averaged
component of physical linear momentum  associated with  a harmonic
plane wave in the direction of its propagation. These conditions
are ensured if
\begin{eqnarray}\label{pmKill}
   \pmKill{K} &=& \frac{\g(K,K)}{|\g(K,K)|}.
\end{eqnarray}
i.e. $\pmKill{K}=1$ if $K$ is spacelike and $\pmKill{K}=-1$ if
$K$ is timelike.

If the spacetime admits a foliation by hypersurfaces $t=$ constant
then (\ref{DTAU}) takes the form adapted to the frame $U$:
\begin{eqnarray*}
    \sd\,\JU{K} + \frac{\dotRU{K}}{\cc} &=& 0.
\end{eqnarray*}

In this spacetime framework, a basic postulate is that the history
of a material continuum interacting with gravity and \EM fields
can be determined from (\ref{DIVT}) and the Maxwell system
(\ref{MAXSYS})
%(Tucker \& Walton 2009):
\begin{eqnarray}\label{MAXSYS}
    dF \eqq 0 \qquadand d\star G \eqq j,
\end{eqnarray}
where the excitation 2-form $G$ depends on the interaction with
the medium and the 3-form electric 4-current $j$ encodes the
electric charge and current source. Such an electric 4-current
describes both (mobile) electric charge and effective (Ohmic)
currents in a conducting medium. To close this system in a
background gravitational field, electromagnetic constitutive
relations relating $G$ and $j$ to $F$ are necessary. These
relations may also depend upon properties of the medium, including
its state of motion.

If one makes the arbitrary split $\TT=\TENSOR^{matter,EM}+
\TENSOR^{EM}$ then in general $\nabla\cdot\TENSOR^{matter,EM} =
-\nabla\cdot \TENSOR^{EM} \neq 0$ so, as in the non-relativistic
modelling situation, any interpretation of various terms in the
decomposition of $\nabla\cdot \TT$ must be understood to be with
respect to a particular splitting. This is of particular relevance
in situations where $\nabla\cdot\TENSOR^{EM}$ contains a coupling
of electromagnetic fields and mass-energy to the bulk local
acceleration field of a medium.

In a fully coupled system that consistently incorporates the
gravitational interactions using Einstein's gravitational field
equations with a symmetric Einstein tensor, (\ref{DIVT}) becomes
an identity. In descriptions with non-dynamic gravitation
(\ref{DIVT}) is part of the coupled system for the remaining
dynamic fields. Since the effects of gravity will be neglected in
the following we assume henceforth a background Minkowski
spacetime and all observers will be inertial with $\nabla {U}=0$.

Consider a material continuum macroscopic model in which $T^{tot}$
contains  the symmetric \EM \SEM tensor
\begin{eqnarray}\label{SM}
    \TENSOR^{SM} &=& \frac{1}{2}\left( \df i_{a}F \tensor i^{a}G + i_{a}G \tensor i^{a}F + \star(F \w \star G) \g \right).
\end{eqnarray}
The excitation $2$-form $G$ in this expression must be specified
in terms of $F$ and other properties of the medium. For a simple
non-dispersive isotropic medium, one has the constitutive relation
%(\ref{DielecCR})
\begin{eqnarray}\label{DielecCR}
    G = \ep{0}\ep{r}i_{V}F \w \wt{V} - \frac{\ep{0}}{\mu_{r}}\star\left(i_{V}\star F \w \wt{V}\right) = \ep{0}\left( \ep{r} - \frac{1}{\mu_{r}} \right)i_{V}F
\w \wt{V} + \frac{\ep{0}}{\mu_{r}} F,
\end{eqnarray}
where $V$ is a unit, timelike 4-velocity field describing the bulk
motion of the medium and $\ep{r}$ and $\mu_{r}$ are relative
permittivity and permeability scalars on spacetime. For a
spatially inhomogeneous medium these will not be constants:
$d\,\ep{r}\neq 0$, $d\,\mu_{r}\neq 0$. Furthermore, if the medium
is accelerating then $\nabla_{V}V\neq 0$. For any Killing vector
field $K$, the tensor $T^{SM}$ gives rise to the Killing drive
3-form:
\begin{eqnarray}\label{TAUKSM}
    \tauK{SM} &=& -\frac{\pmKill{K}}{2}\left( i_{K}G \w \star F - F \w i_{K}\star G\df \right),
\end{eqnarray}
where the forms $F$ and $G$ are required to satisfy the Maxwell
equations (\ref{MAXSYS}). For an uncharged {\it non-conducting}
medium $j=0$. The \SEM tensor (\ref{SM}) is that obtained by
symmetrizing the one proposed by Minkowski to describe \EM
stresses and energy balance in a medium.

%%%%%%%%%%%%%%%%%%%%%%%%%%%%%%%%%%%%%%%%%%%%%%%%%%%%%%%%%%%%%%%%%%%%%%%%%%%%%%%%%%%%%%%%
\section{Covariant Forces and Torques in Media}
%%%%%%%%%%%%%%%%%%%%%%%%%%%%%%%%%%%%%%%%%%%%%%%%%%%%%%%%%%%%%%%%%%%%%%%%%%%%%%%%%%%%%%%%

\begin{figure}[h]
\setlength{\unitlength}{2.5cm}
\begin{center}
    \includegraphics[width=10cm]{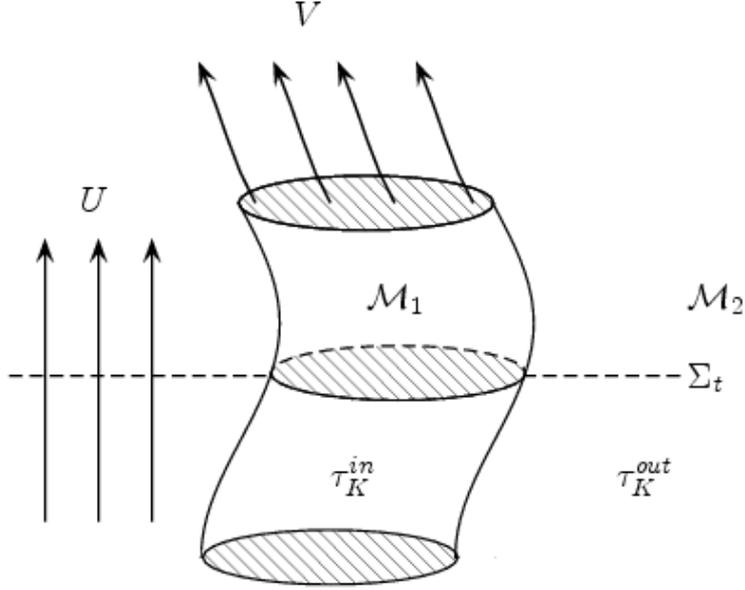}
%%%%%%%%%%%% ------- SG put to the following command --------
\caption{Immersion of a material history in spacetime}\label{pop}
\end{center}
\end{figure}

In general the total \SEM for a medium may contain singularities
and discontinuities in its material properties. To facilitate the
following it proves convenient to restrict to a bounded domain of
spacetime containing (the smooth history of) the medium immersed
in the vacuum (see figure \ref{pop}). Thus we explicitly leave out
of the discussion sources of stress that may arise from
singularities or discontinuities in the medium history. One may
regard  the immersion as the  {\it material  body} and it is
assumed here to have a topologically trivial structure with a
total interior Killing drive form $\tauK{in}$ with interior
support including the smooth boundary of the body. The exterior of
the body in spacetime is assigned a total Killing\footnote{When
the effects of gravity are neglected the interior and exterior
spacetimes admit the same set of Killing vector fields.} drive
$\tauK{out}$ with exterior support excluding the boundary of the
body.

For any observer field $U$, spacetime domain $\M$ and  total
$K$-drive $\tauK{\M}$  on $\M$, one has from (\ref{tau_split})
\begin{eqnarray*}
    J_{K}^{\UUU} \eqq -i_{U}\tauK{\M}  \qquad \text{and} \quad \rho_{K}^{\UUU} \eqq -(i_{U} \star \tauK{\M})\star \wt{U}.
\end{eqnarray*}
If one writes
\begin{eqnarray*}
    \tauK{\M} &=& -\pmKill{K}\cc^{2}\, \wh{\rho}_{me} \g(V,K) \star \wt{V} + \wh{\tau}_{K}^{\M}
\end{eqnarray*}
to describe a medium with bulk 4-velocity $V$ on $\M$ and proper
inertial mass-energy density scalar\footnote[2]{The first term on the
right in the above equation arises from a contribution $\cc^2\,
\wh{\rho}_{me} \wt{V} \otimes \wt{V}$ to the total \SEM tensor.}
$\wh{\rho}_{me}$, given the conservation law
$d(\wh{\rho}_{me}\star \wt{V})=0$, the equation of motion
$d\,\tauK{\M}=0$ becomes
\begin{eqnarray*}
    \pmKill{K}\cc^{2}\,\wh{\rho}_{me} \wt{A}(K) &=& f^{\M}_{K},
\end{eqnarray*}
where $f^{\M}_{K}\equiv -\star \,d\,\wh{\tau}_{K}^{\M}$ and the
4-acceleration vector field of $\M$ is $A\equiv\nabla_{V}{V}$.
Contracting the $\star$ Hodge dual of this equation of motion with
$U$ and integrating the resulting 3-form over any $U$-$orthogonal$
spacelike hypersurface $\SIG$ that intersects the domain $\M$
yields
\begin{eqnarray*}
    \pmKill{K}\int_{\SIG}\, \mu_{U} \wt{A}(K) &=& f_{K}^{U,\M}[\SIG],
\end{eqnarray*}
where the mass-energy 3-form $\mu_{U}\equiv
-\cc^{2}\wh{\rho}_{me}\star \wt{U}$ and the total instantaneous
integrated $K$-drive on $\SIG$ at time $t$ in the $U$ frame is
\begin{eqnarray*}
    f_{K}^{U,\M}[\SIG] &\equiv& \int_{\SIG} \, i_{U}\,d\,\wh{\tau}_{K}^{\M}.
\end{eqnarray*}
Thus for any part $\tauK{\M,j}$ of the total $K$-drive
$\tauK{\M}=\sum_{j} \,\tauK{\M,j}$ on $\M$ we call the 3-form
$-i_{U}\,d\,\tauK{\M,j}$ the $j$-th part of the instantaneous
$K$-drive density 3-form.

Using this notation consider the two regions $\M_{1}=\inn$ and
$\M_2=\outt$ with $\tauK{\inn}$ on $\M_{1}$ describing the
interaction of a medium with \EM fields according to $d\,
\tauK{\inn}=0$ and $\tauK{\outt}$ describing \EM fields in the
vacuum according to $d\, \tauK{\outt}=0$ on $\M_{2}$. The
interface between these regions will be denoted $\sigOneTwo$ regarded
as an immersion in spacetime. In the following the difference of
$J_K^{U,\inn}$ and $J_K^{U,\outt}$ across $\sigOneTwo$ will be
encountered. This discontinuity jump is defined in terms of the
pull-back map $\sigOneTwo^{\star}$ on these forms to $\sigOneTwo$:
\begin{eqnarray*}
    \Delta_{12}(\JUK{}) &=& \sigOneTwo^{\star}(  J_{K}^{U,\inn}  - J_{K}^{U,\outt}).
\end{eqnarray*}
With a split given by $j=\{ matter, EM \}$ write
\begin{eqnarray*}
    \tauK{\inn} &=& \tauK{\inn,matter} + \tauK{\inn,EM}
\end{eqnarray*}
where $\tauK{\inn,matter}$ describes matter without permanent
polarization or magnetization in the absence of external \EM
fields, the equation $d\tauK{\inn}=0$ yields
\begin{eqnarray}\label{LIONN}
    f_{K}^{U,\inn,matter}[\Sigma_{t}]+ f_{K}^{U,\inn,EM}[\Sigma_{t} ] &=& 0.
\end{eqnarray}
As noted in the introduction the \EM constitutive relation may
imply a coupling of the \EM fields in the medium to its
deformation tensor and hence may contribute to its strains. For an
uncharged medium, if there are no \EM fields to polarize the
medium, the second term on the left in (\ref{LIONN}) will be zero.
In this case one may identify it with the $K$-drive on the medium
due to an applied \EM field. Such a $K$-drive will in general give
rise to a non-zero bulk acceleration field $A=\nabla_{V}V$ on
$\M_{1}$. In principle\footnote{This would be difficult in
practise if electromagnetic stresses in a deformable medium led to
significant strains producing a dynamic change in its shape or
volume.} one could maintain any prescribed state of acceleration
$A_{0}$ with $V=W_{0}$ by the addition of a further drive
$f_{K}^{\inn,ext}$ of some nature, provided
\begin{eqnarray*}
    \pmKill{K}\int_{\Sigma_{t}} \, \mu_{U}A_{0}(\wt{K}) &=& f_{K}^{\inn, elast}[\Sigma_{t}] + f_{K}^{\inn,EM}[\Sigma_{t}] + f_{K}^{\inn,ext}[\Sigma_{t} ],
\end{eqnarray*}
where $f_{K}^{\inn, elast}$ isolates possible drives in the medium
due to internal \\%%%% SG -----breaks the line ---------------
non-electromagnetically
induced stresses and
friction with the environment. In particular if the medium is
maintained in a state of uniform rotation in the absence of
interaction with \EM fields by mechanical torques then the
presence of any \EM induced torques can (in principle) be
compensated by additional mechanical torques in order to maintain
a state of uniform rotation. Using the orthogonal decompositions
with respect to $U$ above and the relation $i_{U}d\,
\tauK{\inn,EM}=\sd J_{K}^{\inn,EM} + \Lie_{U}\rho_{K}^{\inn,EM}$,
such an integrated instantaneous external $K$-drive may be
written:
\begin{eqnarray}\label{RESULT}
f_{K}^{U,\inn,ext}[\SIG] = -\int_{\SIG} \, i_{U} d\tauK{\inn,EM} = -\int_{\SIG} \, \Lie_{U}\rho_{K}^{U,\inn,EM} - \int_{\PD{}\SIG} \, J_{K}^{U,\inn,EM},
\end{eqnarray}
where Stokes' theorem has been used in the last term. The precise
nature of this compensating drive will depend of course not only
on the form of $\tauK{\inn,EM}$ but also on the \EM constitutive
relation for the medium.

For {\it unbounded} media the notion of a total integrated drive
may not be meaningful. In these circumstances one may be able to
deduce contributions to $K$-drive pressures from the last term in
(\ref{RESULT}) or by integrating discontinuities of  $J_{K}^{U}$
over surfaces instead of integrating contributions to $i_{U}
d\tauK{}$ over volumes. This follows simply from the bounded
medium case above by writing
\begin{eqnarray*}
    \int_{\PD{}\SIG} \, J_{K}^{U,\inn,EM} &=& \int_{\PD{}\SIG} \, J_{K}^{U,\outt,EM}   + \int_{\PD{}\SIG} \, \Delta_{12}(J_{K}^{U}),
\end{eqnarray*}
where the discontinuity $\Delta_{12}$ is across the boundary of
the history in spacetime. Both integrals now require a knowledge
of the pull-back to the interface of  $J_{K}^{U,\outt}$ in the
vacuum region. However one may relax the condition that the medium
be bounded and assert that the instantaneous $K$-drive exerted on
any {\it finite} spatial volume $\VV$ in an infinite history  having
an interface $S_{t}= \partial\VV$ with the rest of the medium {\it and} the
vacuum, by drive 3-forms that contain discontinuities over part of
$S_{t}$, is given by
\begin{eqnarray}\label{RESULT1}
    f_{K}^{U,\inn,ext}[\VV] &=& -\int_{\VV} \, \Lie_{U}\rho_{K}^{U,\inn,EM} - \int_{S_{t}} \, J_{K}^{U,\outt,EM} - \int_{S_{t}} \, \Delta_{12}(J_{K}^{U}) .
\end{eqnarray}
In static situations the first integral on the right is zero and
in some cases the second integral can be evaluated in terms of
fields or their sources in the vacuum region.\\

%%%%%%%%%%%%%%%%%%%%%%%%%%%%%%%%%%%%%%%%%%%%%%%%%%%%%%%%%%%%%%%%%%%%%%%%%%%%%%%%%%%%%%%%
\section{The Abraham and Symmetrized Minkowski Killing-Drives in Non-Accelerating Media}\label{DRIVES}
%%%%%%%%%%%%%%%%%%%%%%%%%%%%%%%%%%%%%%%%%%%%%%%%%%%%%%%%%%%%%%%%%%%%%%%%%%%%%%%%%%%%%%%%
As noted in the introduction there have been a number of different
proposals for the \EM \SEM tensor. In the tables in Appendix A,
%appendix~\ref{EMTensTable}
the properties of a number of these \cite{MINK,ABR},
%(Minkowski 1908; Abraham 1909),
are displayed using the notation introduced above. For a recent
derivation of these tensors from a variational approach see
\cite{DGT,DGT2}. %(Dereli \textit{et al.} 2006, 2007).
They permit a ready evaluation of the $K$-drives on media  with
bulk $4$-velocity field $V$ and specified constitutive properties.
For convenience, separate entries for the Abraham tensor have been
provided for media at rest relative to a frame $U$ (i.e. $V=U$).
To calculate forces and torques it is only necessary to recognize
that they are defined by particular spacelike Killing vectors $K$.
So in an inertial frame $U$ with $\wt{U}(K)=0$ many entries will
simplify. Separate entries are also given for situations when $K$
is timelike and equal to $U$ since this facilitates computation of
energies and powers. In the calculation of the divergence of the
Abraham \SEM we have made explicit those terms that depend on the
bulk acceleration field $A\equiv \nabla_V\,V$ of the medium. Such
terms are of course absent for media at rest or moving with
arbitrary linear velocity in an inertial frame ($\nabla {U}=0 $).
It is only in special circumstances that there are simple
relations between the Minkowski and Abraham tensors. These
relations arise mainly for simple media {\it at rest} in an
inertial frame. Thus if
\begin{eqnarray}\label{SIMP}
    \Md{U} \eqq \ep{0}\ep{r}\Me{U}, \qquad \Mb{U} \eqq \mu_{0}\mu_{r}\Mh{U}
\end{eqnarray}
in a stationary homogeneous medium (with constant scalars $\ep{r}$ and
$\mu_{r}$), then for all $K$, the $J^{U}_{K}$ are
identical for $T^{SM}, T^{M}$ and $T^{AB}$ (see table in
%appendix~\ref{EMTensTable}).
Appendix A).
However the corresponding $\rhoUK{}$
are all different. In problems with time-harmonic \EM fields the
$\Lie_{U}\rhoUK{}$ have zero averages over time, implying that the
corresponding time-averaged $K$-drives $\TA{f_{K}^{U}}$ are the
same.

In a simple non-accelerating ($A=0$) medium that may be
inhomogeneous there is a simple relation between  $\tauK{A}$ and
$\tauK{SM}$
%(see table in appendix~\ref{EMTensTable})
(see table in Appendix A)
for {\it arbitrary} time-dependent fields.
Given a {\it spacelike translational}
Killing vector $K$ one sees from the table that (since
$\wt{U}(K)=0$)
\begin{eqnarray*}
    \tauK{A} &=& \tauK{SM} - \frac{1}{2}\s{U}(K)\star\wt{U},
\end{eqnarray*}
where the 1-form
\begin{eqnarray*}
    \s{U} &=& \star\left( \frac{1}{\cc}\Me{U} \w \Mh{U} \w \wt{U} - \cc\Md{U} \w \Mb{U} \w \wt{U} \right).
\end{eqnarray*}
Using the simple constitutive relations (\ref{SIMP}) this 1-form
can be expressed in terms of the spatial fields  $\Me{U}, \Mh{U}$
and the form $\wt{U}$
\begin{eqnarray}\label{RWTTT}
    \s{U} &=&  \frac{1}{\cc}\left(1-\man{N}^{2}\right)\star\left( \Me{U} \w \Mh{U} \w \wt{U}\right),
\end{eqnarray}
where $\man{N}\equiv\sqrt{\ep{r}\mu_{r}}$ is the refractive index
of the medium. Thus
\begin{eqnarray*}
    \s{U}(K)\star\wt{U} = -\frac{1}{\cc}\left(1-\man{N}^{2}\right)i_{U}\left( \Me{U} \w \Mh{U} \w \wt{U} \w \wt{K}\right) = \frac{1}{\cc}\left(1-\man{N}^{2}\right)\left( \Me{U} \w \Mh{U} \w \wt{K}\right)
\end{eqnarray*}
and so
\begin{eqnarray}\label{MJT}
    \tauK{A} \eqq \tauK{SM} + \frac{1}{2\cc}\left(\man{N}^{2}-1\right)\left( \Me{U} \w \Mh{U} \w \wt{K}\right).
\end{eqnarray}
By contracting the exterior derivative of this expression with
$U=\frac{1}{\cc}\PD{t}$ (defining a non-accelerating reference
frame) one deduces a relation between the 3-form force density
$\FF^{U,A}_{K}=i_{U}d\tauK{A}$ associated with the Abraham drive
3-form, and $\FF^{U,SM}_{K}=i_{U}d\tauK{SM}$ associated with the
symmetrized Minkowski drive 3-form:
\begin{eqnarray}\label{STAR}
    \FF^{U,A}_{K} \eqq \FF^{U,SM}_{K} + \frac{1}{2\cc^{2}}\Lie_{\PD{t}}\left( \left(\man{N}^{2}-1\right)\Me{U} \w \Mh{U}\right) \w \wt{K},
\end{eqnarray}
using
%(\ref{LIE})
\begin{eqnarray}\label{LIE}
    d\phi \eqq \sd\phi - \wt{U} \w \Lie_{U}\phi \eqq \sd\phi + dt \w \Lie_{\PD{t}}\phi,
\end{eqnarray}
and noting that  $\left(\man{N}^{2}-1\right)\left( \Me{U} \w
\Mh{U} \w \wt{K}\right)$ is a spatial 3-form\footnote[1]{If one
replaces $U$ by $V$ with $\nabla_{V}V\neq 0$ in (\ref{RWTTT}) one
is dealing with accelerating media and the Abraham drive
$\FF^{U,A}_{K}$ then acquires additional terms in (\ref{STAR}).
Such additional terms are often overlooked when considering the
effects of the so called Abraham force (or torque). This may lead
to inconsistencies in processes involving accelerating media.}.
The second term on the right in (\ref{STAR}) has been termed the
Abraham force. In static field situations it is zero and the two
force densities are equal. This is also the situation in the
presence of time-periodic fields if one takes the time-average of
this equation  over a time period. For smooth pulsed fields the
second term will in general yield a finite contribution to a total
force when (\ref{STAR}) is integrated over a finite time interval.

Returning to the constitutive case for an arbitrarily moving
medium (\label{DielecCR}) one can use the results in the previous
section to facilitate a comparison between the Newtonian force
density (\ref{MODEL1}) and that based on the choice with
$\tau_{K}^{\M,EM}= \tauK{SM}$. Thus one must calculate
$i_{U}\,d\,\tauK{SM}$ for a Killing field $K$ that generates {\it
spatial translations} and express the result in terms of the
polarization and magnetization 1-forms $\Mp{U}$ and $\Mm{U}$
respectively. A local constitutive relation between $G$ and $F$
will induce a local constitutive relation between $\Mp{U}$ and
$\Mm{U}$ and the \EM fields $\Me{U}$ and $\Mb{U}$ and the
permittivity and permeability of the medium. Since $d\,F=0$,
\begin{eqnarray*}
    2d\tauK{SM} &=& -\left(di_{K}G \w \star F - i_{K}G \w d\star F - F \w di_{K}\star G\right).
\end{eqnarray*}
Writing  $G=\ep{0}F+\Pi$, one has
\begin{eqnarray*}
    &&di_{K}G \w \star F = \Lie_{K}G \w \star F - i_{K}dG \w \star F = F \w \star \Lie_{K}G - i_{K}d(\ep{0}F + \Pi) \w \star F \\
   && \quad = F \w \Lie_{K}\star G - i_{K}d\Pi \w \star F = F \w i_{K}d\star G + F  \w di_{K}\star G - F \w \star i_{K}d\Pi \\
   && \quad  = F \w i_{K}j + F \w di_{K}\star G - F \w \star i_{K}d\Pi.
\end{eqnarray*}
Hence
\begin{eqnarray*}
    2d\tauK{SM} \eqq -F \w i_{K}j + F \w \star i_{K}d\Pi + i_{K}G \w d\star F.
\end{eqnarray*}
Since
\begin{eqnarray*}
    d\star G =\ep{0}d\star F + d\star\Pi = j, \quad \text{or} \quad d\star F =  \frac{1}{\ep{0}}\left(j - d\star\Pi\right),
\end{eqnarray*}
and
\begin{eqnarray*}
&&i_{K}G \w d\star F = -G \w i_{K}d\star F = - \frac{1}{\ep{0}}G \w
i_{K}\left(j - d\star\Pi\right) \\
&&\quad = -\frac{1}{\ep{0}}G \w i_{K}j + \frac{1}{\ep{0}}G \w i_{K}d\star\Pi
= -F \w i_{K}j - \frac{\Pi}{\ep{0}} \w i_{K}j + G \w i_{K}d\star\frac{\Pi}{\ep{0}},
\end{eqnarray*}
one has
\begin{eqnarray*}
    2d\tauK{SM} &=& -\left(2F + \frac{\Pi}{\ep{0}}\right) \w i_{K}j + F \w \star i_{K}d\Pi + G \w i_{K}d\star\frac{\Pi}{\ep{0}}.
\end{eqnarray*}
This can be decomposed in terms of $\Me{U},\Mb{U},\Mp{U},\Mm{U}$,
and real and induced polarization and magnetization sources
yielding\footnote[2]{
See %appendix~\ref{SMsect}
\ESM for further details of this calculation.}
\begin{eqnarray*}
&&    -2d\tauK{SM} = \mbox{\small $\left( \RU{p}\Me{U}(K) + i_{\wt{\Mb{U}}}\J{p} \w \kappa + \frac{1}{\ep{0}}(\RU{}+\RU{p})\Mp{U}(K)\right) \wedge \wt{U}$} \\
&&\quad \mbox{\small $+\bigg(\mu_{0}i_{\wt{\Mm{U}}}(\J{} + \J{p}) \w \kappa + 2\RU{}\Me{U}(K) + 2i_{\wt{\Mb{U}}}\J{} \w \kappa \,+ \frac{1}{\cc}i_{\wt{\Me{U}}}\J{m} \w \kappa$}
                   \\
&&\quad\qquad \mbox{\small $- \cc\RU{m}\Mb{U}(K)\bigg) \w \wt{U}
 - 2\cc\MB{U} \w i_{K}\RU{} - \frac{1}{\ep{0}\cc}\MM{U}\w
                 i_{K}\RU{}$} \\
&&\quad \mbox{\small $- i_{K}\RU{m} \w \ME{U} + \cc i_{K}\MB{U} \w \RU{p}
                  + \frac{1}{\ep{0}\cc}i_{K}\MM{U} \w \RU{p}$},
\end{eqnarray*}
where $\MM{U}=\#Mm{U}$. Contracting with $U$ yields
\begin{eqnarray}
\begin{split}\label{SMFORCE}
    \FF_{K}^{U,SM} = -i_{U}d\tauK{SM} =&  \mbox{\small $\RU{}\Me{U}(K) + i_{\wt{\Mb{U}}}\J{} \w \kappa  + \frac{1}{2}\RU{p}\Me{U}(K)$} \\& \mbox{\small $+  \frac{1}{2} i_{\wt{\Mb{U}}}\J{p} \w \kappa  + \frac{1}{2\cc}i_{\wt{\Me{U}}}\J{m} \w \kappa -  \frac{1}{2} \cc\RU{m}\Mb{U}(K)$} \\& \mbox{\small $+ \frac{1}{2\ep{0}}(\RU{}+\RU{p})\Mp{U}(K) +  \frac{1}{2} \mu_{0}i_{\wt{\Mm{U}}}(\J{} + \J{p}) \w \kappa$}.
\end{split}
\end{eqnarray}
The first two terms on the right constitute the Lorentz force
density due to any real charge $\RU{}$ and electric current $\J{}$
in the medium. The third and fourth terms constitute the Lorentz
force density  due to induced polarization  charge $\RU{p}$ and
electric current $\J{p}$. These are precisely the force densities
that arise in the non-relativistic electrically neutral smoothed
model (\ref{MODEL1}) above based on the motion of point charges.
The fifth and sixth terms constitute the Lorentz force density
due to magnetization charge $\RU{m}$ and magnetization current
$\J{m}$. These do not arise in the Newtonian model (\ref{MODEL1})
above since the point charges in that model were not endowed with
intrinsic magnetic
moments.

%%%%%%%%%%%%%%%%%%%%%%%%%%%%%%%%%%%%%%%%%%%%%%%%%%%%%%%%%%%%%%%%%%%%%%%%%%%%%%%%%%%%%%%%%
\section{Conclusions}
%%%%%%%%%%%%%%%%%%%%%%%%%%%%%%%%%%%%%%%%%%%%%%%%%%%%%%%%%%%%%%%%%%%%%%%%%%%%%%%%%%%%%%%%%
%It is concluded
%that if a normally incident plane harmonic vacuum wave excites a
%plane harmonic mode in a (uniformly rotating or
%stationary) medium, subject to the Minkowski constitutive relations, the induced time-averaged \EM torque will be
%zero. Thus one cannot discriminate between the Minkowski and Abraham
%\SEM tensors by exciting such modes in a simple medium in
%this manner.
%This motivates our exploration in paper II of the
% effects of incident plane harmonic vacuum waves on plane slabs of inhomogeneous uniformly rotating media.

The general theory of drive-forms has been developed starting from
the vanishing  divergence of a total  \SEM in an arbitrary
spacetime using the language of exterior systems. It has been
shown that a decomposition of a drive 3-form,  relative to a unit
timelike vector field, on a spacetime with sufficient  Killing
vectors  yields forms that can be associated with force and torque
densities in continuous media. For material subject to the
electromagnetic constitutive relations proposed by Minkowski it
has also been shown how the computation of  the \EM force density
on a non-accelerating  polarizable perfectly insulating medium  in
an \EM field can be effectively carried out in terms of a
particular split of the total \SEM into parts  describing its
inertial and  \EM properties. In paper II this theory is applied
to homogeneous and inhomogeneous  dielectric media where it is
argued that a means of discriminating between a split into the \EM
\SEM tensors proposed by Abraham and Minkowski (and possibly
others) can be explored experimentally using rotating media.

%%%%%%%%%%%%%%%%%%%%%%%%%%%%%%%%%%%%%%%%%%%%%%%%%%%%%%%%%%%%%%%%%%%%%%%%%%%%%%%%%%%%%%%%%
\section{Acknowledgements}
%%%%%%%%%%%%%%%%%%%%%%%%%%%%%%%%%%%%%%%%%%%%%%%%%%%%%%%%%%%%%%%%%%%%%%%%%%%%%%%%%%%%%%%%%
The authors are grateful to the Cockcroft Institute, the Alpha-X
project, STFC and EPSRC (EP/E001831/1) for financial support for
this research.
%%%%%%%%%%%%%%%%%%%%%%%%%%%%%%%%%%%%%%%%%%%%%%%%%%%%%%%%%%%%%%%%%%%%%%%%%%%%%%%%%%%%%%%%%

%%%%%%%%%%%%%%%%%%%%%%%%%%%%%%%%%%%%%%%%%%%%%%%%%%%%%%%%%%%%%%%%%%%%%%%%%%%%%%%%%%%%%%%%%

\begin{landscape}
\appendix{}
\section{Summary of Electromagnetic Stress-Energy-Momentum Tensors}
\footnotesize
%%%%%%%%%%%%%%%%%%%%%%%%%%%%%%%%%%%
\renewcommand\arraystretch{2.5}
%%%%%%%%%%%%%%%%%%%%%%%%%%%%%%%%%%%
\begin{tabular}{|p{1cm}|p{7cm}|p{10.4cm}|}
    \hline
    & \centering MINKOWSKI & \hspace{3.6cm} SYMMETRIZED MINKOWSKI  \\
    \hline
    \centering$\TENSOR^{EM}$ & \hspace{0.1cm} $i_{a}F \tensor i^{a}G + \frac{1}{2}\star(F \w \star G) g $ & \hspace{0.1cm} $\TENSOR^{SM} = \frac{1}{2}\left( \df i_{a}F \tensor i^{a}G + i_{a}G \tensor i^{a}F + \star(F \w \star G) g \right)$    \\

    \hline
    \centering$\tauK{EM}$ & \hspace{0.1cm} $-\frac{\pmKill{K}}{2} \left( \df i_{K}F \w \star G - F \w i_{K}\star G\right)$ & \hspace{0.1cm} $-\frac{\pmKill{K}}{2}\left(i_{K}G \w \star F - F \w i_{K}\star G\df\right)$ \\

    \hline
    \centering$\nabla\cdot\TENSOR^{EM}$ & \hspace{0.1cm}$ \frac{1}{2}\df d\star( F \w \star G ) - (\nabla_{X_{a}}G)(\wt{i^{a}F}) - G(\wt{\nabla \cdot F})  $ & \hspace{0.1cm}$\frac{1}{2}\left(\df d\star( F \w \star G ) - (\nabla_{X_{a}}G)(\wt{i^{a}F}) - (\nabla_{X_{a}}F)(\wt{i^{a}G}) - G(\wt{\nabla \cdot F}) - F(\wt{\nabla \cdot G}) \right)$  \\

    \hline
    \centering$\JU{K}$ & \hspace{-0.5cm}
%%%%%%%%%%%%%%%%%%%%%%%%%
\renewcommand\arraystretch{2.8}
%%%%%%%%%%%%%%%%%%%%%
\begin{tabular}{rl}
                                                & $-\pmKill{K}\left[\Me{U}(K)\MD{U} + \Mh{U}(K)\MB{U} - \frac{1}{\cc}\wt{U}(K)\Me{U} \w \Mh{U}\right.$ \\

                                                & $\left.- \frac{1}{2}\left( \Me{U}\!\cdot\Md{U} + \Mb{U}\!\cdot\Mh{U}\right)\kappa\right]$
                                             \end{tabular}\vspace{0.3cm}
                       &
%%%%%%%%%%%%%%%%%%%%%%%%%%%%%%%%%%%%
\renewcommand\arraystretch{2.8}
%%%%%%%%%%%%%%%%%%%%%%%%%%%%%%%%%%%5
\begin{tabular}{rl}
                                                & $-\frac{\pmKill{K}}{2}\left[\Me{U}(K)\MD{U}+ \Md{U}(K)\ME{U} + \Mh{U}(K)\MB{U} +\Mb{U}(K)\MH{U} \right.$ \\

                                                & $\left.- \left(\Me{U}\!\cdot\Md{U} +  \Mb{U}\!\cdot\Mh{U} \right)\kappa - \wt{U}(K)\left( \frac{1}{\cc}\Me{U} \w \Mh{U} + \cc\Md{U} \w \Mb{U} \right)\right]$
                                            \end{tabular} \\
    \hline
    \centering$\RU{K}$ &
%%%%%%%%%%%%%%%%%%%%%%%%%%%%%%%%%
 \renewcommand\arraystretch{2.8}
%%%%%%%%%%%%%%%%%%%%%%%%%%%%%%%%
\begin{tabular}{rl}
                                                & $\pmKill{K}\left[\cc\Md{U} \w \Mb{U} \w \wt{K}^{\perp}\right.$ \\

                                                & $\left. + \frac{1}{2}\wt{U}(K)\left(\Me{U}\!\cdot\Md{U} + \Mb{U}\!\cdot\Mh{U}\right)\#1\right]$

                                             \end{tabular}\vspace{0.3cm} & \hspace{0.1cm} $ \frac{\pmKill{K}}{2}\left[\left(\frac{1}{\cc}\Me{U} \w \Mh{U}+ \cc\Md{U} \w \Mb{U}\right)\w \wt{K}^{\perp} + \wt{U}(K)\left(\Me{U}\!\cdot\Md{U} + \Mb{U}\!\cdot\Mh{U} \right)\#1\right]$ \\

    \hline
    \centering$\JU{U}$ & \hspace{0.1cm}$\frac{1}{\cc}\Me{U} \w \Mh{U}$ & \hspace{0.1cm} $\frac{1}{2}\left( \frac{1}{\cc}\Me{U} \w \Mh{U} + \cc\Md{U} \w \Mb{U} \right)$\\

    \hline
    \centering$\RU{U}$ & \hspace{0.1cm}$\frac{1}{2}\left(\Me{U}\!\cdot\Md{U} + \Mb{U}\!\cdot\Mh{U}\right)\#1$ & \hspace{0.1cm}$\frac{1}{2}\left(\Me{U}\!\cdot\Md{U} + \Mb{U}\!\cdot\Mh{U}\right)\#1$ \\

    \hline
\end{tabular}

%%%%%%%%%%%%%%%%%%%%%%%%%%%%%%%%%
 \renewcommand\arraystretch{2.8}
%%%%%%%%%%%%%%%%%%%%%%%%%%%%%%%%
\begin{tabular}{|p{0.6cm}|p{9.4cm}|p{8.5cm}|}

    \hline
    & \centering ABRAHAM $(U\neq V)$ &  \hspace{3.4cm} ABRAHAM $(U=V)$  \\
    \hline
    \centering$\TENSOR^{EM}$ & \hspace{0.1cm} $\frac{1}{2}\left( \df i_{a}F \tensor i^{a}G + i_{a}G \tensor i^{a}F + \star(F \w \star G) g - \s{V} \tensor \wt{V} - \wt{V} \tensor \s{V}\right)$ & \hspace{0.1cm}$\frac{1}{2}\left( \df i_{a}F \tensor i^{a}G + i_{a}G \tensor i^{a}F + \star(F \w \star G) g - \s{U} \tensor \wt{U} - \wt{U} \tensor \s{U}\right)$  \\

    \hline
    \centering$\tauK{EM}$ & \hspace{0.1cm} $-\frac{\pmKill{K}}{2}\left(i_{K}G \w \star F - F \w i_{K}\star G - \s{V}(K)\star \wt{V} - \wt{V}(K)\star \s{V}  \df\right)$ & \hspace{0.1cm} $-\frac{\pmKill{K}}{2}\left(i_{K}G \w \star F - F \w i_{K}\star G - \s{U}(K)\star \wt{U} - \wt{U}(K)\star \s{U} \df\right)$ \\

    \hline
    \centering$\nabla\cdot\TENSOR^{EM}$ & \hspace{0.1cm} $\nabla\cdot \TENSOR^{SM} - \frac{1}{2}\left((\nabla\cdot\s{V})\wt{V}  + (\nabla\cdot\wt{V})\s{V} \df\right) - \Psi(V) - F(\wt{G(\man{A}})) + G(\wt{F(\man{A}}))$ & \hspace{0.1cm} $\nabla\cdot \TENSOR^{SM} - \frac{1}{2}\left((\nabla\cdot\s{U})\wt{U}  + (\nabla\cdot\wt{U})\s{U}  \df\right) - \Psi(U)$ \\

    \hline
    \centering$\JU{K}$ &\hspace{-0.4cm}
%%%%%%%%%%%%%%%%%%%%%%%%%%%%%%%%
 \renewcommand\arraystretch{2.8}
%%%%%%%%%%%%%%%%%%%%%%%%%%%%%%%%
\begin{tabular}{rl}
                                                & $-\frac{\pmKill{K}}{2}\left[ \Me{U}(K)\MD{U}+ \Md{U}(K)\ME{U} + \Mh{U}(K)\MB{U} +\Mb{U}(K)\MH{U} \right.$ \\

                                                & $- \left(\Me{U}\!\cdot\Md{U} +  \Mb{U}\!\cdot\Mh{U} \right)\kappa - \wt{U}(K)\left( \frac{1}{\cc}\Me{U} \w \Mh{U} + \cc\Md{U} \w \Mb{U} \right)$ \\

                                                & $\left.-i_{U}\left( \wt{K} \w i_{V}\star\s{V}\right) + 2\wt{V}(K)i_{U}\star\s{V}\right]$

                                            \end{tabular}\vspace{0.3cm}
              & \hspace{-0.4cm}$
%%%%%%%%%%%%%%%%%%%%%%%%%%%%%%%%
\renewcommand\arraystretch{2.8}
%%%%%%%%%%%%%%%%%%%%%%%%%%%%%%%
\begin{tabular}{rl}
                                                & $-\pmKill{K}\left[\frac{1}{2}\left( \Me{U}(K)\MD{U}+ \Md{U}(K)\ME{U} + \Mh{U}(K)\MB{U} +\Mb{U}(K)\MH{U} \right)\right.$ \\

                                                & $\left.- \frac{1}{2}\left(\Me{U}\!\cdot\Md{U} +  \Mb{U}\!\cdot\Mh{U} \right)\kappa - \frac{1}{\cc}\wt{U}(K)\Me{U} \w \Mh{U} \right]$  \\

                                            \end{tabular} $\\
    \hline
    \centering$\RU{K}$ & \hspace{-0.6cm}
%%%%%%%%%%%%%%%%%%%%%%%%%%%%%%%%%%
\renewcommand\arraystretch{2.8}
%%%%%%%%%%%%%%%%%%%%%%%%%%%%%%%%%%
\begin{tabular}{rl}
                                                & $ \frac{\pmKill{K}}{2}\left[\left(\frac{1}{\cc}\Me{U} \w \Mh{U}+ \cc\Md{U} \w \Mb{U}\right)\w \wt{K}^{\perp} - \wt{U}(V)\wt{U}(K)\star\s{V} \right.$ \\

                                                & $ \left. + \wt{U}(K)\left(\Me{U}\!\cdot\Md{U} + \Mb{U}\!\cdot\Mh{U} \right)\#1 + \wt{U}(V)i_{U}\star\s{V} \w \wt{K}  + \wt{V}(K)\Pi_{U}\star\s{V}\right]$

                                            \end{tabular}\vspace{0.3cm}
             & \hspace{-0.4cm}
%%%%%%%%%%%%%%%%%%%%%%%%%%%%%%
 \renewcommand\arraystretch{2.8}
%%%%%%%%%%%%%%%%%%%%%%%%%%%%%%
\begin{tabular}{rl}
                                                & $ \pmKill{K}\left[\frac{1}{\cc}\Me{U} \w \Mh{U} \w \wt{K}^{\perp} + \frac{1}{2}\wt{U}(K)\left(\Me{U}\!\cdot\Md{U} + \Mb{U}\!\cdot\Mh{U} \right)\#1\right]$

                                            \end{tabular} \\
    \hline
    \centering$\JU{U}$ &
%%%%%%%%%%%%%%%%%%%%%%%%%%%%%%%
 \renewcommand\arraystretch{2.8}
%%%%%%%%%%%%%%%%%%%%%%%%%%%%%%%
\begin{tabular}{rl}
                                                & $\frac{1}{2}\left( \frac{1}{\cc}\Me{U} \w \Mh{U} + \cc\Md{U} \w \Mb{U} \right) + \frac{1}{2}\Pi_{U}i_{V}\star\s{V} + \wt{U}(V)i_{U}\star\s{V}$

                                            \end{tabular}
              & \hspace{0.1cm} $\frac{1}{\cc}\Me{U} \w \Mh{U}$ \\
    \hline
    \centering$\RU{U}$ & \hspace{0.1cm}
    $\frac{1}{2}\left(\Me{U}\!\cdot\Md{U} + \Mb{U}\!\cdot\Mh{U}\right)\#1
      - \wt{U}(V)\Pi_{U}\star \s{V}$ & \hspace{0.1cm}
      $\frac{1}{2}\left(\Me{U}\!\cdot\Md{U} + \Mb{U}\!\cdot\Mh{U}\right)\#1$ \\

    \hline
\end{tabular}

\end{landscape}
\normalsize
%%%%%%%%%%%%%%%%%%%%%%%%%%%%%%%%%%%%%%%%%%%%%%%%%%%%%%%%%%%%%%%%%%%%%%%%%%%%%%%%%%%%%%%%%
%\section{Summary of Electromagnetic Stress-Energy-Momentum  Tensors and their Drive Form
%Properties}\label{EMTensTable}
%%%%%%%%%%%%%%%%%%%%%%%%%%%%%%%%%%%%%%%%%%%%%%%%%%%%%%%%%%%%%%%%%%%%%%%%%%%%%%%%%%%%%%%%%
In these tables we have introduced the 1-form:
\begin{eqnarray*}
    \s{V} &=& \star (i_{V}F \w i_{V}\star G \w \wt{V} - i_{V}G \w i_{V}\star F \w \wt{V})  \\
           &=& \star \left(\frac{1}{\cc}\Me{V} \w \Mh{V} \w \wt{V} - \cc\Md{V} \w \Mb{V} \w \wt{V}\right),
\end{eqnarray*}
where the spatial fields $\Me{V}, \Mb{V}, \Md{V}, \Mh{V} $ are
defined by the orthogonal splits of $F$ and $G$ relative to the
4-velocity field $V$ of the medium. Furthermore, we have defined
$\MH{U}=\#\Mh{U}$. The vector fields $U$ and $V$ on spacetime are
timelike, unit, normalized with a metric $\g$ of signature
$(-1,+1,+1,+1)$ so that $\wt{U}(U)=\wt{V}(V)=-1$, and describe the
state of motion of the observer and medium respectively. It is
also convenient to introduce, in terms of the vector field $U$ and
any vector field $Y$ on spacetime the projection $Y^{\perp}= Y +
\wt{U}(Y)U$, which induces the projection operator on $p$-forms
$\Pi_{U} = \mathbb{I} + \wt{U} \w i_{U}$. The spatial Hodge map
$\#$  with respect  to $U$ is defined by $\star 1 = \wt{U} \w \#
1$, and maps spatial $p$-forms to spatial $(3-p)$-forms. For any
Killing vector field $K$ on spacetime (i.e. $\Lie_{K}\g=0$) we
denote the 2-form $\#\wt{K}$ by $\kappa$. If $K$ is a Killing
vector then so is $\xi_{0}\,K$ where $\xi_{0}$ is any constant.
Hence the physical dimensions of quantities that depend on the
Killing vector field $\xi_{0}\,K$ will depend on the physical
dimensions of $\xi_{0}$.

%The 3-form $\tauK{}$  associated with a symmetric
%stress-energy-momentum tensor $\TENSOR$  is constructed by contraction with a Killing
%vector field $K$ followed by taking the spacetime Hodge dual:
%\begin{eqnarray*}
%   \tauK{} &=& -\pmKill{K}\star(\TENSOR(K)),
%\end{eqnarray*}
%where
%$\pmKill{K}=1$ if $K$ is spacelike and $-1$ if $K$ is timelike.
%The orthogonal split of $\tauK{}$ with respect to $U$ is defined
%by
%\begin{eqnarray*}
%   \tauK{} &=& \JU{K} \w \wt{U} + \RU{K} \\
%  \text{with} \qquad \JU{K} &\equiv& -i_{U}\tauK{} \qquad \text{and} \qquad \RU{K} \;\;\equiv\;\; -(i_{U}\star\tauK{})\star\wt{U}.
%\end{eqnarray*}
In the table we have isolated those terms in the divergence that
depend {\it explicitly} on the bulk medium acceleration
$A\equiv\nabla_{V}V$. Among the remaining terms there is a set
that can be expressed in terms of a (2,0)-tensor field $\Psi$ on
spacetime defined, for any vector fields $X,Y$, by
$\Psi(X,Y)=(\Psi(X))(Y)$ where
\begin{eqnarray*}
    \Psi(X) &\equiv& \mbox{\small $(\nabla_{X}F)(\wt{i_{X}G}) + F(\wt{i_{X}\nabla_{X}G}) - (\nabla_{X}G)(\wt{i_{X}F}) - G(\wt{i_{X}\nabla_{X}F}) - \frac{1}{2}(\Lie_{X}F)(\wt{i_{X}G})$} \\
            & & \mbox{\small $- \frac{1}{2}F((\Lie_{X}\ginv)(i_{X}G)) - \frac{1}{2}F(\wt{i_{X}\Lie_{X}G}) + \frac{1}{2}(\Lie_{X}G)(\wt{i_{X}F}) + \frac{1}{2}G((\Lie_{X}\ginv)(i_{X}F))$}  \\
            & & \mbox{\small $+ \frac{1}{2}G(\wt{i_{X}\Lie_{X}F}) + \frac{1}{2}(\Lie_{X}\g)(\wt{F(\wt{i_{X}G})})  - \frac{1}{2}(\Lie_{X}\g)(\wt{G(\wt{i_{X}F})})$}
\end{eqnarray*}
with $F(Y)\equiv i_{Y}F$ and $G(Y) \equiv i_{Y}G$. If $\{X_a \}$ denotes an orthonormal frame with dual cobasis $\{ e^b\}$ one has
${\g}^{-1}= \eta^{ab}\, X_a \otimes X_b ,\quad  {\g}=\eta_{ab} \, e^a \otimes e^b$
where $\eta^{ab}$ and $\eta_{ab}$ are both diagonal matrices with $\eta^{00}=\eta_{00}=-1$   and $\eta_{ij}=\eta^{ij}=\delta_{ij}$ for $i,j=1,2,3$.  The contraction operator  $i_{X_a}$ is abbreviated $i_a$ and $i^a\equiv \eta^{ab}\,i_b$.
The spacetime Hodge map is denoted $\star$ and the canonical
4-form on spacetime is $\star 1 =  e^0 \wedge e^1 \wedge e^2 \wedge e^3$ in an orthonormal basis and is $\sqrt{\vert{  \text{det} }\,\g
\vert} \,d^{4}x$ in any coordinate system $ \{ x^a \}  $.

%%%%%%%%%%%%%%%%%%%%%%%%%%%%%%%%%%%%%%%%%%%%%%%%%%%%%%%%%%%%%%%%%%%%%%%%%%%%%%%%%%%%%%%%
\section{Notation}
%%%%%%%%%%%%%%%%%%%%%%%%%%%%%%%%%%%%%%%%%%%%%%%%%%%%%%%%%%%%%%%%%%%%%%%%%%%%%%%%%%%%%%%%
The natural mathematical language to discuss the differential
properties of tensor fields on spacetime and their relation to
integrals over material domains is in terms of differential forms
and their associated exterior calculus \cite{RWT}.
%(Benn \& Tucker 1988).
In this section a brief summary is given of the relevant notation
used in subsequent sections. A key concept throughout involves the
role of the spacetime metric tensor field and possible isometries
that it may possess. A spacetime metric tensor field $\g$ is a
symmetric bilinear form on spacetime that can always be
represented in a local cobasis of differential 1-forms $\{e^{a}\}$
as
\begin{eqnarray}
    \g &=& -e^{0}\tensor e^{0} +  e^{1}\tensor e^{1} + e^{2}\tensor e^{2} + e^{3}\tensor e^{3}.
\end{eqnarray}
If $\{X_{b}\}$ is the dual local basis of vector fields on
spacetime defined so that $e^{a}(X_{b})=\delta^{a}_{b}$
$(a,b,=0,1,2,3)$ one has the induced contravariant metric
\begin{eqnarray}
    \ginv &=& -X_{0}\tensor X_{0} +  X_{1}\tensor X_{1} + X_{2}\tensor X_{2} + X_{3}\tensor X_{3}.
\end{eqnarray}
If $\alpha$ is any given 1-form, one has an associated vector
field $\wt{\alpha}$ defined so that $\beta(\wt{\alpha})=
\ginv(\alpha,\beta)$ for all 1-forms $\beta$.  Since $\ginv$ is
symmetric this will be abbreviated $\wt{\alpha}=\ginv(\alpha)$. In
a similar way if $X$ is any given vector field one has an
associated 1-form $\wt{X}= \g(X)$.

In the $\g$-orthonormal basis $\{e^{a}\}$ one has a canonical
local 4-form denoted $\star 1$ and defined to be $e^{0}\w e^{1} \w
e^{2} \w e^{3}$. The Hodge map $\star$ induced by $\star 1$ maps
$p$-forms to $(4-p)$-forms on spacetime
%\cite{RWT}.
(Benn \& Tucker 1988; see also Appendix A of paper I). The metric
also uniquely defines the covariant derivative $\nabla_{X}$ with
respect to any vector field $X$. This has the property
$\nabla_{X}\g=0$ and $\nabla_{X}Y - \nabla_{Y}X = [X,Y]$ for all
vector fields $X,Y$. In this expression $[X,Y]$ denotes the
commutator bracket. While $\nabla_{X}$ has a type-preserving
action on any tensor field the exterior derivative $d$ is defined
to act only on antisymmetric tensor fields (differential forms)
and has the property $d \circ d=0$. Contraction of any $p$-form
$\beta$ with $X$ is denoted $i_{X}\beta$. The interior operator
$i_{X}$ is a graded derivation defined so that
\begin{eqnarray}
    i_{X}(\alpha \w \beta) &=& (i_{X}\alpha) \w \beta + (-1)^{p}\, \alpha \w i_{X}\beta,
\end{eqnarray}
for any $p$-form $\alpha$ and $q$-form $\beta$. If $p=1$, one
defines $i_{X}\alpha=\alpha(X)$, and if $p=0$, $i_{X}\alpha=0$,
for all vector fields $X$. One has the useful relations
\begin{eqnarray*}
    \star\star\Phi &=& (-1)^{p+1}\Phi \\
    i_{X}\star\Phi &=& \star ( \Phi \w \wt{X} ) \\
    \Lie_{X}\Phi &=& i_{X}d\Phi + di_{X}\Phi,
\end{eqnarray*}
for any $p$-form $\Phi$ on spacetime where $\Lie_{X}$ denotes Lie
differentiation with respect to $X$.

The spacetime divergence operator $\nabla\cdot$ takes a simple
form if one uses the $\g$-orthonormal basis above. Acting on a
symmetric covariant tensor $T$ it defines by contraction  on the
first argument the $1$-form
\begin{eqnarray} \label{DIVDEF}
    \nabla\cdot T &=& \sum_{a=0}^{3}(\nabla_{X_{a}}\, T) ( X^{a},-)
\end{eqnarray}
where $\{X^{a}\}=\{-X_{0},X_{1},X_{2},X_{3}\}$. For a symmetric
tensor $T$ it is sufficient to write the right hand side as
$(\nabla_{X_{a}}\, T) ( X^{a})$. A local spacetime isometry with
respect to $\g$ is a local diffeomorphism that preserves this
metric. A vector field $K$ that generates such a diffeomorphism is
called a Killing vector field and it satisfies $\Lie_{K}\g=0$ in
terms of the operation of Lie differentiation with respect to $K$.
The operators $\star, d, i_{X}, \nabla_{X}, \Lie_{X}$ offer a
powerful computational tool-kit when working with differential
forms.

For any {\it smooth} $p$-form $\Phi$ in a bounded regular region
$\M$ of a manifold one can express the integral of $d\Phi$ over
$\M$ in terms of the integral of $\Phi$ over the boundary
$\PD{}\M$ of $\M$:
\begin{eqnarray}
    \int_{\M}\, d\Phi &=& \int_{\PD{}\M}\, \Phi .
\end{eqnarray}
This is a statement of Stokes' theorem for $p$-forms.

The Gibbs calculus of vector fields in 3-dimensional Euclidean
space is readily exposed by correspondences induced by the
exterior operations above. A space of 3-dimensions may be
considered as a particular hypersurface in spacetime. If the
metric above induces the metric
\begin{eqnarray}
    {\mathbf \g} &=& e^{1}\tensor e^{1} + e^{2}\tensor e^{2} + e^{3}\tensor e^{3}
\end{eqnarray}
on this hypersurface it is Euclidean and one may introduce the
Euclidean canonical form $\# 1= e^{1}\w e^{2} \w e^{3}$ by
restriction. Since spacetime is assumed  time-oriented one may
employ a future-pointing timelike unit vector field $U$ on
spacetime (with $\g(U,U)=-1$ and $\wt{U}=e^{0}$) to fix a coherent
orientation by relating $\#1$ to $\star 1$ by
\begin{eqnarray}
    \star 1 &=& \wt{U} \w \#1 .
\end{eqnarray}
If the hypersurface is given as $t=$ constant for some time
coordinate $t$, an inertial frame exists in Minkowski spacetime
with $U=\frac{1}{\cc}\pdiff{}{t}=-X_{0}$ such that $\nabla\,U=0$.
Throughout this article the constant $\cc$ denotes the speed of
light in the vacuum. Any $p$-form $\alpha$ on spacetime is termed
spatial with respect to such a $U$ if $i_{U}\alpha=0$. An over-dot
will denote (Lie) differentiation with respect to the coordinate
$t$  so
\begin{eqnarray}
    \dot{\alpha} &\equiv& \cc\Lie_{U}\alpha = \Lie_{\PD{t}}\alpha.
\end{eqnarray}
On a Euclidean hypersurface in spacetime, exterior differentiation
of spatial $p$-forms $\phi$ is denoted  $\sd\phi$ such that
\begin{eqnarray}\label{LIE}
    d\phi &=& \sd\phi - \wt{U} \w \Lie_{U}\phi = \sd\phi + dt \w \Lie_{\PD{t}}\phi.
\end{eqnarray}
One has the following relations
\begin{eqnarray*}
    \#1 &=& -i_{U}\star 1 = -\star \wt{U} \\
    \# \# \phi &=& \phi
\end{eqnarray*}
for all spatial $p$-forms $\phi$. Furthermore if $\Evec, \Ewec$
denote Euclidean vector fields in the Gibbs notation corresponding
to the vector fields $\wt{v}, \wt{w}$ for some spatial 1-forms $v,
w$ then
\begin{eqnarray}
    \label{EDIV} \text{div } \Evec \qquad &&\text{corresponds to} \qquad \wt{\# \sd \# v}\; ; \\
\nonumber && \\
    \label{CURL} \text{curl } \Evec \qquad &&\text{corresponds to} \qquad \widetilde{\# \sd v}\; ; \\
\nonumber && \\
    \label{VPRODUCT} {\Evec \times \Ewec} \qquad &&\text{corresponds to} \qquad \widetilde{ \# ( v \w w )}\; ;\\
\nonumber && \\
    \label{GRAD} \text{grad } \psi \qquad &&\text{corresponds to} \qquad \wt{\sd \psi},
\end{eqnarray}
for any scalar $\psi$  field on spacetime.\\

%%%%%%%%%%%%%%%%%%%%%%%%%%%%%%%%%%%%%%%%%%%%%%%%%%%%%%%%%%%%%%%%%%%%%%%%%%%%%%%%%%%%%%%%%
\section{Electromagnetic Fields on Spacetime}
%%%%%%%%%%%%%%%%%%%%%%%%%%%%%%%%%%%%%%%%%%%%%%%%%%%%%%%%%%%%%%%%%%%%%%%%%%%%%%%%%%%%%%%%%
We suppose that a polarizable material continuum is given in terms
of a set of piecewise smooth material properties that determine
its  interaction with classical gravitational and \EM fields. The
classical macroscopic Maxwell system for the electromagnetic
2-form $F$ on spacetime can be written as
%\cite{GCM8}:
(Tucker \& Walton 2009):
\begin{eqnarray}\label{MAXSYS}
    dF = 0 \qquadand d\star G = j,
\end{eqnarray}
where the excitation 2-form $G$ depends on the interaction with
the medium  and the 3-form electric 4-current $j$ encodes the
electric charge and current source\footnote{
    All electromagnetic tensors in this article have
    dimensions constructed from the SI dimensions $[M], [L], [T], [Q]$ where $[Q]$
    has the unit of the Coulomb in this system. We adopt $[\g]=[L^{2}],
    [G]=[j]=[Q],\,[F]=\frac{[Q]}{[\ep{0}]}$ where the permittivity of free space
    $\ep{0}$ has the dimensions $[ Q^{2} T^{2} M^{-1} L^{-3}]$ and
    $\cc=\frac{1}{\sqrt{\ep{0}\mu_{0}}}$ denotes the speed of light in vacuo.
    Note that the operators $d$ and $\nabla$  preserve the physical dimensions of tensor fields but with $[\g ]=[L^{2}]$, for $p$-forms $\alpha$ in $4$ dimensions,
    one has $[\star \alpha]=[\alpha] [L^{4-2p}]$.}.
Such an electric 4-current describes both (mobile) electric charge
and effective (Ohmic) currents in a conducting medium. To close
this system in a background gravitational field, {\it
electromagnetic constitutive relations} relating $G$ and $j$ to
$F$ are necessary.

The history of a particular observer field in spacetime is
associated with an arbitrary {\it unit} future-pointing timelike
4-velocity vector field $U$. The field $U$ may be used to describe
an {\it observer frame} on spacetime and its integral curves model
{\it idealized observers}. An orthogonal decomposition of $F$ with
respect to any observer field $U$ gives rise to a pair of {\it
spatial} 1-forms on spacetime. The 1-form spatial {\it electric
field} $\Me{U}$ and 1-form spatial {\it magnetic induction field}
$\Mb{U}$ associated with $F$ are defined with respect to an
observer field $U$ by
\begin{eqnarray}\label{intro_e_b}
    \Me{U} = i_{U}F \qquadand \cc\Mb{U} = i_{U} \star F.
\end{eqnarray}
Since $\g(U,U)=-1$ and $i_{U}\Me{U}=i_{U}\Mb{U}=0$:
\begin{eqnarray}\label{intro_F}
    F &=& \Me{U}\w\wt{U} - \star\,(\cc\Mb{U}\w \wt{U}).
\end{eqnarray}
Likewise the 1-form spatial {\it displacement field} $\Md{U}$ and
the 1-form spatial {\it magnetic field} $\Mh{U}$ associated with
$G$ are defined with respect to $U$ by
\begin{eqnarray}\label{Media_d_h}
    \Md{U} = i_{U}G  \qquadand \frac{\Mh{U}}{\cc} = i_{U}\star G,
\end{eqnarray}
so
\begin{eqnarray}\label{Media_G}
    G &=& \Md{U}\w \wt{U} - \star\left( \frac{\Mh{U}}{\cc}\w \wt{U} \right),
\end{eqnarray}
with $i_{U}\Md{U}=i_{U}\Mh{U}=0$. At the history of any sharp
interface between different media, given as the piecewise smooth
(non-null) spacetime hypersurface $f=0$, the system of Maxwell
equations is supplemented by interface conditions on the fields
$F$ and $G$
\begin{eqnarray}\label{MaxwellBC}
\begin{split}
    \left.\df[F]\right|_{f=0} \w df &= 0 \\
    \left.\df[\star G]\right|_{f=0} \w df &= j_{s},
\end{split}
\end{eqnarray}
where $[H]$ denotes the discontinuity in the field $H$ across the
hypersurface. The 3-form $j_{s}$ on the hypersurface is non zero
if it supports a real current 3-form there.\\

%%%%%%%%%%%%%%%%%%%%%%%%%%%%%%%%%%%%%%%%%%%%%%%%%%%%%%%%%%%%%%%%%%%%%%%%%%%%%%%%%%%%%%%%%
\section{Time-Dependent Maxwell Systems in 3-Space}
%%%%%%%%%%%%%%%%%%%%%%%%%%%%%%%%%%%%%%%%%%%%%%%%%%%%%%%%%%%%%%%%%%%%%%%%%%%%%%%%%%%%%%%%%
The spacetime description above is natural for the Maxwell system
since it makes no reference to any particular frame in spacetime.
However to make contact with descriptions in particular frames or
non-relativistic formulations a reduction in terms of frame
dependent fields becomes mandatory. The spacetime Maxwell system
can now be reduced to a family of parameterized exterior systems
on $\real^{3}$. Each member is an exterior system involving forms
on $\real^{3}$ depending parametrically on some time coordinate
$t$ associated with $U$. Let the $(3+1)$ split of the electric
4-current 3-form with respect to a foliation  of spacetime by
spacelike hypersurfaces with constant $t$  be
\begin{eqnarray}\label{decompJ}
    j &=& \frac{\J{}}{\cc} \w \wt{U} + \RU{},
\end{eqnarray}
with $i_{U}\J{}=i_{U}\RU{}=0$ and $\RU{}=\hatRU{}\# 1$, where
$\J{},\RU{}$ are the spatial electric current density 2-form and
spatial electric charge density 3-form respectively. The
differential operator $\sd$ on spacetime forms is adapted to those
spacetimes (such as Minkowski spacetime where gravity is absent)
that can be foliated by hypersurfaces with constant coordinate $t$
where $U=\frac{1}{\cc}\PD{t}$. Then, from (\ref{MAXSYS}),
\begin{eqnarray}\label{dj}
    dj &=& 0
\end{eqnarray}
yields
\begin{eqnarray}\label{cont}
    \sd\man{J}^{U} + \dot{\rho}^{U} &=& 0.
\end{eqnarray}
 The $(3+1)$ split of the
spacetime covariant Maxwell equations (\ref{MAXSYS}) with respect
to $\wt{U}=-\cc dt$ becomes
\begin{eqnarray}
     \label{M1} \sd\,\Me{U} &=& -\dotMB{U} \\
     \label{M2} \sd\,\MB{U} &=& 0 \\
     \label{M3} \sd\,\Mh{U} &=& \J{} + \dotMD{U} \\
     \label{M4} \sd\,\MD{U} &=& \RU{}
\end{eqnarray}
where $\MD{U}=\#\Md{U}$ and $\MB{U}=\#\Mb{U}$. All $p$-forms
($p>0$) in these equations are independent of $dt$, but have
components that may depend parametrically on $t$.\\

%%%%%%%%%%%%%%%%%%%%%%%%%%%%%%%%%%%%%%%%%%%%%%%%%%%%%%%%%%%%%%%%%%%%%%%%%%%%%%%%%%%%%%%%%
\section{Electromagnetic Constitutive Relations}
%%%%%%%%%%%%%%%%%%%%%%%%%%%%%%%%%%%%%%%%%%%%%%%%%%%%%%%%%%%%%%%%%%%%%%%%%%%%%%%%%%%%%%%%%
The two 2-forms $F$ and $G$ in the macroscopic Maxwell equations
on spacetime are fundamentally related by smoothing the
microscopic sources of  the \EM  fields in the medium. In many
circumstances one then relies on phenomenological relations for
closure relations. In such relations the excitation form $G$ is in
general a functional (possibly non-local in spacetime) of the
Maxwell form $F$, its covariant differentials, thermodynamic
properties, deformation properties and the state of motion of the
medium:
\begin{eqnarray*}
    G &=& \man{Z}[F, \nabla\,F\ldots].
\end{eqnarray*}
Such a functional may induce  non-linear and non-local relations
between $\Md{U}, \Mh{U}$ and $\Me{U}, \Mb{U}$. Electrostriction
and magnetostriction arise from the dependence of $\man{Z}$ on the
deformation tensor of the medium and its covariant derivatives.
For general {\it linear continua},  a knowledge of a collection of
{\it constitutive tensor fields} $Z^{(r)}$ on spacetime may
suffice so that
\begin{eqnarray*}
    G &=& \sum_{r=0}^{N}Z^{(r)}[\nabla^{r}F,\ldots].
\end{eqnarray*}
In idealized (non-dispersive) {\it simple continua}, one adopts
the idealized {\it local} relation
\begin{eqnarray*}
    G &=& Z(F),
\end{eqnarray*}
for some degree 4 constitutive tensor field $Z$, parameterized by
scalars that depend on the medium. In the vacuum $G=\ep{0}F$ where
$\ep{0}$ is the constant permittivity of the vacuum. Regular {\it
lossless, non-conducting, linear isotropic media} can be
described by a bulk 4-velocity field $V$ of the medium, a real
relative permittivity scalar field $\ep{r} > 0$ and a real
relative permeability scalar field $\mu_{r} > 0 $. In this case,
the structure of the tensor $Z$ follows from
\begin{eqnarray}\label{DielecCR}
\begin{split}
    G &\eqq \ep{0}\ep{r}i_{V}F \w \wt{V} - \frac{\ep{0}}{\mu_{r}}\star\left(i_{V}\star F \w \wt{V}\right) \\
      &\eqq \ep{0}\left( \ep{r} - \frac{1}{\mu_{r}} \right)i_{V}F \w \wt{V} + \frac{\ep{0}}{\mu_{r}} F.
\end{split}
\end{eqnarray}
For inhomogeneous media the relative permittivity and permeability
scalars $\ep{r}$ and $\mu_{r}$ will not be constants. In a general
frame $U$ comoving with the medium ($U=V$), (\ref{DielecCR})
yields
\begin{eqnarray}\label{dielecComCR}
    \Md{V} = \ep{0}\ep{r}\Me{V} \qquadand \Mh{V} = (\mu_{0}\mu_{r})^{-1}\Mb{V},
\end{eqnarray}
which are the familiar closure relations for simple (idealized)
electrically neutral isotropic non-dispersive polarizable media.\\

%%%%%%%%%%%%%%%%%%%%%%%%%%%%%%%%%%%%%%%%%%%%%%%%%%%%%%%%%%%%%%%%%%%%%%%%%%%%%%%%%%%%%%%%%%%%%%%%%%%%
\subsection{Minkowski Constitutive Relations for Moving Media}
%%%%%%%%%%%%%%%%%%%%%%%%%%%%%%%%%%%%%%%%%%%%%%%%%%%%%%%%%%%%%%%%%%%%%%%%%%%%%%%%%%%%%%%%%%%%%%%%%%%%
The above algebraic constitutive relation (\ref{DielecCR})
involves the bulk 4-velocity field of a simple medium. It is
straightforward to find the induced relations between the fields
$\{\Me{U},\Mb{U},\Md{U},\Mh{U}\}$  relative to an observer in a
frame $U \neq V$. From (\ref{intro_F}) and (\ref{Media_G}), one
has
\begin{eqnarray*}
    \Me{V} &=& i_{V}F = \Me{U}(V)\wt{U} - \wt{U}(V)\Me{U} - \star\left( \cc\Mb{U} \w \wt{U} \w \wt{V}\right) \\
    \cc\Mb{V}&=& i_{V}\star F = \star(\Me{U} \w \wt{U} \w \wt{V}) +  \cc\Mb{U}(V)\wt{U} - \cc\wt{U}(V)\Mb{U} \\
    \Md{V} &=& i_{V}G = \Md{U}(V)\wt{U} - \wt{U}(V)\Md{U} - \star\left( \frac{\Mh{U}}{\cc} \w \wt{U} \w \wt{V}\right) \\
    \frac{\Mh{V}}{\cc} &=& i_{V}\star G = \star\left(\Md{U} \w \wt{U} \w \wt{V}\right) + \frac{\Mh{U}(V)}{\cc}\wt{U} - \frac{\wt{U}(V)}{\cc}\Mh{U}.
\end{eqnarray*}
Inserting these in the constitutive relations (\ref{dielecComCR})
yields
\begin{eqnarray*}
\begin{split}%\label{transCR}
    \mbox{\small $\Md{U}(V)\wt{U} - \wt{U}(V)\Md{U} - \star\left( \frac{\Mh{U}}{\cc} \w \wt{U} \w \wt{V} \right)$} &= \mbox{\small $\ep{}\left[\df \Me{U}(V)\wt{U} - \wt{U}(V)\Me{U} - \star\left( \cc\Mb{U} \w \wt{U} \w \wt{V}\right)\right]$} \\
    \mbox{\small $\cc\star\left(\Md{U} \w \wt{U} \w \wt{V}\right) + \Mh{U}(V)\wt{U} - \wt{U}(V)\Mh{U}$} &= \mbox{\small $\mu^{-1}\!\left[\!\df\star\left(\frac{\Me{U}}{\cc} \w \wt{U} \w \wt{V}\right) +  \Mb{U}(V)\wt{U} - \wt{U}(V)\Mb{U} \right]$},
\end{split}
\end{eqnarray*}
where $\ep{}=\ep{r}\ep{0},\mu=\mu_{r}\mu_{0}$ in terms of the
relative permittivity scalar $\ep{r}$ and relative permeability
$\mu_{r}$. Contracting with $U$ gives the relations
\begin{eqnarray}
\begin{split}\label{UconCR}
    \Md{U}(V) &= \ep{}\Me{U}(V) \\
    \Mh{U}(V) &= \mu^{-1}\Mb{U}(V),
\end{split}
\end{eqnarray}
which yields
\begin{eqnarray}
\begin{split}\label{transCR2}
    \wt{U}(V)\Md{U} + \star\left( \frac{\Mh{U}}{\cc} \w \wt{U} \w \wt{V} \right) &= \ep{}\left[\df\wt{U}(V)\Me{U} + \star\left( c\Mb{U} \w \wt{U} \w \wt{V}\right)\right] \\
    \wt{U}(V)\Mb{U} - \star\left(\frac{\Me{U}}{\cc} \w \wt{U} \w \wt{V}\right)  &= \mu\left[\df\wt{U}(V)\Mh{U} - \cc\star\left(\Md{U} \w \wt{U} \w \wt{V}\right)\right].
\end{split}
\end{eqnarray}
A laboratory Minkowski {\it inertial frame}  is described by
$U=\frac{1}{\cc}\PD{t}$  in  inertial Cartesian coordinates
$\{t,x^1,x^2,x^3\}  $  with $\{e^0=-\cc dt, e^1=d x^1,e^2= d
x^2,e^3= d x^3\} $   and, relative to $U$, the medium 4-velocity
$V$ has the orthogonal decomposition:
\begin{eqnarray*}
    V &=& \gamma\left( U + \frac{\V}{\cc} \right),
\end{eqnarray*}
where the spatial field  $\V$ is the Newtonian velocity
field\footnote{In inertial Cartesian coordinates $\V =
\sum_{j=1}^{3} v^{j}(t,x^{1},x^{2},x^{3}) \pdiff{}{x^{j}}$.} of
the medium relative to $U$,  $\gamma^{-1} \equiv \sqrt{(1-
\frac{\nu^2}{\cc^2})} $ and $\nu^{2}\equiv\g(\V,\V)$. With these
definitions it follows that:
\begin{eqnarray*}
    \g(U,V) = -\gamma \qquad \text{and} \qquad  \wt{U} \w \wt{V} = \frac{\gamma}{\cc}\wt{U} \w  \Vt.
\end{eqnarray*}
Using these in (\ref{transCR2}) and rearranging yields
\begin{eqnarray}
\begin{split}\label{transCR3}
    \Md{U} - \star\left( \wt{U} \w \Vt \w \frac{\Mh{U}}{\cc^{2}}  \right) &= \ep{}\left[\df\Me{U} - \star\left( \wt{U} \w \Vt \w \Mb{U} \right)\right] \\
    \Mb{U} + \star\left(\wt{U} \w \Vt \w  \frac{\Me{U}}{\cc^{2}} \right)  &= \mu\left[\df\Mh{U} + \star\left( \wt{U} \w \Vt \w \Md{U} \right)\right],
\end{split}
\end{eqnarray}
or using the spatial Hodge map  $\#$ defined by $U$:
\begin{eqnarray}
\begin{split}\label{MinkCR}
    \Md{U} + \#\left( \Vt \w \frac{\Mh{U}}{\cc^{2}}  \right) &= \ep{}\left[\df\Me{U} + \#\left( \Vt \w \Mb{U} \right)\right] \\
    \Mb{U} - \#\left( \Vt \w  \frac{\Me{U}}{\cc^{2}} \right)  &= \mu\left[\df\Mh{U} - \#\left( \Vt \w \Md{U} \right)\right].
\end{split}
\end{eqnarray}
These are the constitutive relations first written by Minkowski.
For some purposes it is useful to decouple these expressions and
express $\Me{U}$ and $\Mh{U}$ directly in terms of $\Md{U},\Mb{U}$
and $\V$.\\

%%%%%%%%%%%%%%%%%%%%%%%%%%%%%%%%%%%%%%%%%%%%%%%%%%%%%%%%%%%%%%%%%%%%%%%%%%%%%%%%%%%%%%%%%%%%%%%%%%%%
\subsection{Minkowski Constitutive Relations for  $\Me{U}(\Md{U},\Mb{U},\V),\Mh{U}(\Md{U},\Mb{U},\V)$}
%%%%%%%%%%%%%%%%%%%%%%%%%%%%%%%%%%%%%%%%%%%%%%%%%%%%%%%%%%%%%%%%%%%%%%%%%%%%%%%%%%%%%%%%%%%%%%%%%%%%
Taking the exterior product of (\ref{MinkCR}) with $\V$ yields,
\begin{eqnarray*}
    \mu\Vt \w \Mh{U} &=& \Vt \w \Mb{U} - \Vt \w \#\left( \Vt \w  \frac{\Me{U}}{\cc^{2}} \right) + \mu\Vt \w \#\left( \Vt \w \Md{U} \right).
\end{eqnarray*}
For any spatial 1-form  $\alpha^{U}$ with respect to $U$ one has:
\begin{eqnarray}
    \nonumber \Vt \w \#\left( \Vt \w \alpha^{U}\right) &=& -\Vt \w \star\left(\wt{U} \w \Vt \w \alpha^{U} \right) = -\star i_{\V}\left( \wt{U} \w \Vt \w \alpha^{U} \right) \\
    \nonumber                                 &=& \nu^{2}\star \left( \wt{U} \w \alpha^{U} \right) - \alpha^{U}(\V)\star\left( \wt{U} \w \Vt \right) \\
    \label{spatident}                         &=& -\nu^{2}\#\alpha^{U} + \alpha^{U}(\V)\#\Vt.
\end{eqnarray}
Thus
\begin{eqnarray*}
    \mu\Vt \w \Mh{U} &=& \Vt \w \Mb{U} + \frac{\nu^{2}}{\cc^{2}}\#\Me{U} - \frac{\Me{U}(\V)}{\cc^{2}}\#\Vt  - \mu\nu^{2}\#\Md{U} + \mu\Md{U}(\V)\#\Vt \\
                     &=& \Vt \w \Mb{U} + \frac{\nu^{2}}{\cc^{2}}\#\Me{U} - \mu\nu^{2}\#\Md{U} - \frac{1}{\ep{}\cc^{2}}(1-\ep{}\mu \cc^{2})\Md{U}(\V)\#\Vt,
\end{eqnarray*}
using the identity $\Md{U}(\V)=\ep{}\Me{U}(\V)$, obtained by
contracting (\ref{MinkCR}) with $\V$. Since $\#\#=1$
\begin{eqnarray*}
    \mu\#\left(\Vt \w \Mh{U}\right) &=& \#\left(\Vt \w \Mb{U}\right) + \frac{\nu^{2}}{\cc^{2}}\Me{U} - \mu\nu^{2}\Md{U} - \frac{1}{\ep{}\cc^{2}}(1-\ep{}\mu \cc^{2})\Md{U}(\V)\Vt.
\end{eqnarray*}
Substituting this into the first relation of (\ref{MinkCR}) yields
\begin{eqnarray*}
    \left(\ep{}\mu \cc^{2} - \frac{\nu^{2}}{\cc^{2}}\right)\Me{U} &=& \mu \cc^{2}\left(1-\frac{\nu^{2}}{\cc^{2}}\right)\Md{U} - \left(\ep{}\mu \cc^{2} -1\right)\left( \#\left(\Vt \w \Mb{U}\right) - \frac{\Md{U}(\V)}{\ep{}\cc^{2}}\Vt\right).
\end{eqnarray*}
Since
\begin{eqnarray*}
    \frac{1}{\ep{}\mu} &=& \frac{\cc^{2}}{\man{N}^{2}},
\end{eqnarray*}
where $\man{N}^{2}=\ep{r}\mu_{r}$ is the square of the medium
refractive index scalar, this may be written
\begin{eqnarray*}
    \left(\man{N}^{2} - \frac{\nu^{2}}{\cc^{2}}\right)\Me{U} &=& \frac{\man{N}^{2}}{\ep{}}\left(1-\frac{\nu^{2}}{\cc^{2}}\right)\Md{U} - \left(\man{N}^{2}-1\right)\left( \#\left(\Vt \w \Mb{U}\right) - \frac{\Md{U}(\V)}{\ep{}\cc^{2}}\Vt \right).
\end{eqnarray*}
Furthermore from (\ref{MinkCR})
\begin{eqnarray*}
    \ep{}\#\left(\Vt \w \Me{U}\right) &=& \#\left(\Vt \w \Md{U}\right) + \left( \frac{\Mh{U}(\V)}{\cc^{2}}\Vt - \frac{\nu^{2}}{\cc^{2}}\Mh{U} \right) - \ep{}\left(\Mb{U}(\V)\Vt - \nu^{2}\Mb{U}\right) \\
                             &=& \#\left(\Vt \w \Md{U}\right)  - \frac{\nu^{2}}{\cc^{2}}\Mh{U} + \ep{} \nu^{2}\Mb{U} - \frac{1}{\mu \cc^{2}}\left(\ep{}\mu \cc^{2} -1 \right)\Mb{U}(\V)\Vt ,
\end{eqnarray*}
using $\Mb{U}(\V)=\mu\Mh{U}(\V)$, obtained by contracting
(\ref{MinkCR}) with $\V$. Substituting this into the second
relation of (\ref{MinkCR}) yields
\begin{eqnarray*}
    \left(\man{N}^{2} - \frac{\nu^{2}}{\cc^{2}}\right)\Mh{U}  &=& \frac{\man{N}^{2}}{\mu}\left( 1- \frac{\nu^{2}}{\cc^{2}}\right)\Mb{U} - \left(\man{N}^{2}-1 \right)\left(-\#\left( \Vt \w \Md{U} \right) -  \frac{\Mb{U}(\V)}{\mu \cc^{2}}\Vt \right).
\end{eqnarray*}
Thus, the constitutive relations (\ref{DielecCR}) can also be
written
\begin{eqnarray*}
\begin{split}\label{DBeh}
    \left(\man{N}^{2} - \frac{\nu^{2}}{\cc^{2}}\right)\Me{U} & = \frac{\man{N}^{2}}{\ep{}}\left(1-\frac{\nu^{2}}{\cc^{2}}\right)\Md{U} + \left(\man{N}^{2}-1\right)\left( -\#\left(\Vt \w \Mb{U}\right) - \frac{\Md{U}(\V)}{\ep{}\cc^{2}}\Vt \right) \\
    \left(\man{N}^{2} - \frac{\nu^{2}}{\cc^{2}}\right)\Mh{U}  & = \frac{\man{N}^{2}}{\mu}\left( 1- \frac{\nu^{2}}{\cc^{2}}\right)\Mb{U} - \left(\man{N}^{2}-1 \right)\left(-\#\left( \Vt \w \Md{U} \right) -  \frac{\Mb{U}(\V)}{\mu \cc^{2}}\Vt \right).
\end{split}
\end{eqnarray*}
In the non-relativistic limit (to first order in
$\frac{\nu}{\cc}$) these constitutive relations become
\begin{eqnarray}
\begin{split}\label{FOmink}
    \Me{U}  &\;\;\approx\;\;  \frac{\Md{U}}{\ep{0}\ep{r}} - \left(1-\frac{1}{\ep{r}\mu_{r}}\right) \#\left(\Vt \w \Mb{U}\right)  \\
    \Mh{U}  &\;\;\approx\;\; \frac{\Mb{U}}{\mu_{0}\mu_{r}} + \left(1 - \frac{1}{\ep{r}\mu_{r}}\right) \#\left(\Vt \w \Md{U}\right).\\
\end{split}
\end{eqnarray}

%%%%%%%%%%%%%%%%%%%%%%%%%%%%%%%%%%%%%%%%%%%%%%%%%%%%%%%%%%%%%%%%%%%%%%%%%%%%%%%%%%%%%%%%%
\section{Polarization and Magnetization}
%%%%%%%%%%%%%%%%%%%%%%%%%%%%%%%%%%%%%%%%%%%%%%%%%%%%%%%%%%%%%%%%%%%%%%%%%%%%%%%%%%%%%%%%%
The polarization 2-form $\Pi$ in spacetime is defined by
\begin{eqnarray}
    \Pi &=& G - \ep{0}F.\label{POL}
\end{eqnarray}
The second  macroscopic Maxwell equation may then be written
\begin{eqnarray*}
    \ep{0}d\star F &=& j - d\star\Pi = j + j_{p},
\end{eqnarray*}
where
\begin{eqnarray}\label{Polcurrent3}
    j_{p} &=& -d\star\Pi
\end{eqnarray}
will be called the electric polarization current 3-form. With
respect to {\it any} observer frame $U$ its orthogonal
decomposition is
\begin{eqnarray}\label{Pi}
    \Pi &=& \Mp{U} \w \wt{U} - \star\left(\frac{\Mm{U}}{\cc} \w \wt{U} \right) = \Mp{U} \w \wt{U} - \frac{1}{\cc}\MM{U} ,
\end{eqnarray}
where $\MM{U}=\#\Mm{U}$ and we call
\begin{eqnarray*}
    \Mp{U} = i_{U}\Pi \qquad \text{and} \qquad \frac{\Mm{U}}{\cc} = i_{U}\star \Pi
\end{eqnarray*}
the spatial polarization 1-form and magnetization 1-form
respectively relative to $U$. The Hodge dual of $\Pi$ has the
decomposition
\begin{eqnarray}\label{starPi}
    \star\Pi &=& \star(\Mp{U} \w \wt{U}) + \frac{\Mm{U}}{\cc} \w \wt{U}  = \MP{U} + \frac{1}{\cc}\Mm{U} \w \wt{U} ,
\end{eqnarray}
where $\MP{U}=\#\Mp{U}$. From (\ref{Media_G}),   (\ref{intro_F}),
(\ref{POL}) and  (\ref{Pi}) it follows
\begin{eqnarray}\label{coPolarCR}
    \Md{U} = \ep{0}\Me{U} + \Mp{U}  \qquad &\text{and}& \qquad \Mh{} = \mu_{0}^{-1}\Mb{U} + \Mm{U}.
\end{eqnarray}
From (\ref{Polcurrent3}), (\ref{Pi})   and (\ref{LIE}) one finds
\begin{eqnarray*}
    j_{p} &=& -d\MP{U} - \frac{d\Mm{U}}{\cc} \w \wt{U} = -\sd\MP{U} + \wt{U} \w \Lie_{U}\MP{U} - \frac{\sd\Mm{U}}{\cc} \w \wt{U}  \\
          &=& -\sd\MP{U} + \frac{1}{\cc}\left( \cc\Lie_{U}\MP{U} -\sd\Mm{U} \right) \w \wt{U}  \\
          &=& -\sd\MP{U} + \frac{1}{\cc}\left( \dotMP{U} - \sd\Mm{U} \right) \w \wt{U}.
\end{eqnarray*}
Writing the orthogonal decomposition of $j_{p}$ with respect to
$U$ as
\begin{eqnarray*}
    j_{p} &=& \frac{\J{p}}{\cc} \w \wt{U} + \RU{p},
\end{eqnarray*}
it follows that
\begin{eqnarray*}\label{polarJrho}
    \frac{\J{p}}{\cc} = -i_{U}j_{p} = \frac{1}{\cc}\left( \dotMP{U} - \sd\Mm{U}\right) \quad \text{and} \quad \RU{p} = -(i_{U}\star j_{p})\star\wt{U} = -\sd\MP{U}.
\end{eqnarray*}
In the frame $U$, $\J{p}$ and $\RU{p}$ denote the induced electric
polarization current density spatial 2-form and induced
polarization charge density spatial 3-form respectively. In a
similar manner
\begin{eqnarray*}
    d\Pi &=& d\Mp{U} \w \wt{U} - \frac{1}{\cc}d\MM{U} = \sd\Mp{U} \w \wt{U} - \frac{1}{\cc}\sd\MM{U} + \frac{1}{\cc}\wt{U} \w \Lie_{U}\MM{U} \\
         &=& \frac{1}{\cc}\left( \cc\sd\Mp{U} + \Lie_{U}\MM{U}\right) \w \wt{U} - \frac{1}{\cc}\sd\MM{U}.
\end{eqnarray*}
with the orthogonal decomposition $j_{m}=d\Pi=\frac{\J{m}}{\cc} \w
\wt{U} + \RU{m}$ where
\begin{eqnarray*}\label{MagJrho}
    \frac{\J{m}}{\cc} = -i_{U}j_{m} = \frac{1}{\cc}\left(\cc\sd\Mp{U} + \frac{1}{\cc}\dotMM{U}\!\right) \quad \text{and} \quad \RU{m} = -(i_{U}\star j_{m})\star\wt{U} = -\frac{1}{\cc}\sd\MM{U}
\end{eqnarray*}
denote the induced magnetization charge current density spatial
2-form and induced magnetization charge density spatial 3-form
respectively in terms of $\Mp{U}$ and $\MM{U}$.\\

%%%%%%%%%%%%%%%%%%%%%%%%%%%%%%%%%%%%%%%%%%%%%%%%%%%%%%%%%%%%%%%%%%%%%%%%%%%%%%%%%%%%%%%%%%%%%%%%%%%%%%%%%%%%%%%%%%%%%%%%%%%%%%%%%

%   Appendices and Tables

%%%%%%%%%%%%%%%%%%%%%%%%%%%%%%%%%%%%%%%%%%%%%%%%%%%%%%%%%%%%%%%%%%%%%%%%%%%%%%%%%%%%%%%%%%%%%%%%%%%%%%%%%%%%%%%%%%%%%%%%%%%%%%%%%%%%

%%%%%%%%%%%%%%%%%%%%%%%%%%%%%%%%%%%%%%%%%%%%%%%%%%%%%%%%%%%%%%%%%%%%%%%%%%%%%%%%%%%%%%%%%
\section{Derivation of the Model Force Density 3-form}\label{SchwingerSect}
%%%%%%%%%%%%%%%%%%%%%%%%%%%%%%%%%%%%%%%%%%%%%%%%%%%%%%%%%%%%%%%%%%%%%%%%%%%%%%%%%%%%%%%%%
The key steps in deriving the expression for the force density
3-form in \S 2 of paper I are outlined (equation 2.14), with the
aid of the identities
\begin{eqnarray}
    \label{spatIdent1} \#\alpha \w \beta &=& \#i_{\wt{\beta}}\alpha, \\
    \label{spatIdent2} \Lie_{\wt{\beta}}\alpha &=& i_{\wt{\beta}}\sd\alpha + \sd i_{\wt{\beta}}\alpha \qquad (\text{Cartan's identity}) \\
    \label{spatIdent3} \sd(\alpha \w \#\beta) &=& 0
\end{eqnarray}
for any spatial 1-forms $\alpha,\beta$. From equation (2.8) of \S
2, paper I, we have
\begin{eqnarray}\label{MODEL}
    \FF_{K} &=&  \frac{1}{2}\left[ \sd\left( \Me{U}\!\cdot\Mp{U} + \Mb{U}\!\cdot\Mm{U} \right) + \# \Lie_{\PD{t}}\left( \Mp{U} \w \Mb{U}\right)  \right] \w \kappa,
\end{eqnarray}
with $\kappa=\#\wt{K}$. The first term may be expressed
\begin{eqnarray*}
    \sd\left(\Me{U}\!\cdot\Mp{U}\right) &=& \sd i_{\wt{\Mp{U}}}\Me{U} = \Lie_{\wt{\Mp{U}}}\Me{U} - i_{\wt{\Mp{U}}}\sd\Me{U} \\
                               &=& \Lie_{\wt{\Mp{U}}}\Me{U} - \#\#i_{\wt{\Mp{U}}}\sd\Me{U} = \Lie_{\wt{\Mp{U}}}\Me{U} - \#(\# \sd\Me{U} \w \Mp{U}),
\end{eqnarray*}
using (\ref{spatIdent1}) and (\ref{spatIdent2}). The last term can
be written
\begin{eqnarray*}
    \#\Lie_{\PD{t}}\left( \Mp{U} \w \Mb{U}\right) &=& \#\left( \dotMp{U} \w \Mb{U} + \Mp{U} \w \dotMb{U}\right) = \#\left( \dotMp{U} \w \Mb{U} - \Mp{U} \w \#\sd\Me{U}\right)  \\
                                         &=& \#( \dotMp{U} \w \Mb{U}) + \#(\#\sd\Me{U} \w  \Mp{U}),
\end{eqnarray*}
using (\ref{M1}). Finally,  the second term is
\begin{eqnarray*}
    \sd\left(\Mb{U}\!\cdot\Mm{U}\right) &=& \sd i_{\wt{\Mb{U}}}\Mm{U} = \Lie_{\wt{\Mb{U}}}\Mm{U} - i_{\wt{\Mb{U}}}\sd\Mm{U} \\
                               &=& \Lie_{\wt{\Mb{U}}}\Mm{U} - \#\#i_{\wt{\Mb{U}}}\sd\Mm{U} = \Lie_{\wt{\Mb{U}}}\Mm{U} - \#(\# \sd\Mm{U} \w \Mb{U}),
\end{eqnarray*}
using (\ref{spatIdent1}) and (\ref{spatIdent2}). From these
expressions
\begin{eqnarray*}
    2\,\man{F}_{K} &=& \left[ \Lie_{\wt{\Mp{U}}}\Me{U} + \Lie_{\wt{\Mb{U}}}\Mm{U} + \#\left( (\dotMp{U} - \# \sd\Mm{U}) \w \Mb{U}\right)  \right] \w \kappa \\
                &=& \left[ \Lie_{\wt{\Mp{U}}}\Me{U} + \Lie_{\wt{\Mb{U}}}\Mm{U} + \#\left( \#\J{p} \w \Mb{U}\right)  \right] \w \kappa \\
                &=& \left[ \Lie_{\wt{\Mp{U}}}\Me{U} + \Lie_{\wt{\Mb{U}}}\Mm{U} + i_{\wt{\Mb{U}}}\J{p}  \right] \w \kappa,
\end{eqnarray*}
using (\ref{polarJrho}) and (\ref{spatIdent1}).  Using Cartan's
identity in the first term here gives
\begin{eqnarray*}
    \Lie_{\wt{\Mp{U}}}\Me{U} \w \kappa &=& \Lie_{\wt{\Mp{U}}}\left( \Me{U} \w \kappa \right) - \Me{U} \w \Lie_{\wt{\Mp{U}}}\kappa \\
                                       &=& \sd i_{\wt{\Mp{U}}}\left( \Me{U} \w \kappa \right) - \Me{U} \w \sd i_{\wt{\Mp{U}}}\kappa - \Me{U} \w i_{\wt{\Mp{U}}}\sd\kappa \\
                                       &=& \sd i_{\wt{\Mp{U}}}\left( \Me{U}(K) \# 1 \right) - \Me{U} \w \sd\#(\wt{K} \w \Mp{U})  \\
                                       &=& \sd\left( \Me{U}(K)\MP{U} \right) + \Me{U} \w \sd\#(\Mp{U} \w \wt{K})  ,
\end{eqnarray*}
using (\ref{spatIdent2}) and (\ref{spatIdent3}) and the fact that
$K=\frac{\PD{}}{\PD{} {x^i}}$ and so $d\kappa=0$. In the last term
one has
\begin{eqnarray}\label{EKPident}
\begin{split}
    \Me{U} \w \sd\#(\Mp{U} \w \wt{K})  &= \Me{U} \w \sd i_{K}\MP{U}  = \Me{U} \w \Lie_{K}\MP{U} - \Me{U} \w i_{K}\sd\MP{U} \\
                                       &= \Me{U} \w \#\Lie_{K}\Mp{U} - \Me{U}(K)\sd\MP{U} \\
                                       &= \Lie_{K}\Mp{U} \w \ME{U} + \RU{p}\Me{U}(K),
\end{split}
\end{eqnarray}
since $K$ is Killing and using (\ref{polarJrho}). This yields
\begin{eqnarray*}
    \Lie_{\wt{\Mp{U}}}\Me{U} \w \kappa &=& \RU{p}\Me{U}(K) + \Lie_{K}\Mp{U} \w \ME{U} + \sd\left( \Me{U}(K)\MP{U} \right)
\end{eqnarray*}
and so the force drive $3$-form  becomes
\begin{eqnarray*}
    \man{F}_{K} &=& \frac{1}{2}\left[\RU{p}\Me{U}(K) + i_{\wt{\Mb{U}}}\J{p} \w \kappa + \Lie_{K}\Mp{U} \w \ME{U}  + \Lie_{\wt{\Mb{U}}}\Mm{U} \w \kappa + \sd\left( \Me{U}(K)\MP{U} \right)\right] .
\end{eqnarray*}

%%%%%%%%%%%%%%%%%%%%%%%%%%%%%%%%%%%%%%%%%%%%%%%%%%%%%%%%%%%%%%%%%%%%%%%%%%%%%%%%%%%%%%%%%
\section{Orthogonal Decompositions used in Deriving the Drive 3-form Associated with the Symmetrized Minkowski Electromagnetic Stress-Energy-Momentum Tensor}\label{SMsect}
%%%%%%%%%%%%%%%%%%%%%%%%%%%%%%%%%%%%%%%%%%%%%%%%%%%%%%%%%%%%%%%%%%%%%%%%%%%%%%%%%%%%%%%%%
In this section, the key steps used in deriving equation (4.6) of
\S 4, paper I are outlined. For $\wt{U}(K)=0$, one calculates
\begin{eqnarray*}
   \frac{\Pi}{\ep{0}} \w i_{K}j &=& \Pi \w \left(\frac{1}{\cc}i_{K}\J{} \w \wt{U} +  i_{K}\RU{} \right) \\
                                &=& \frac{1}{\ep{0}}\left( \Mp{U} \w \wt{U} - \frac{1}{\cc}\MM{U} \right) \w \left(\frac{1}{\cc}i_{K}\J{} \w \wt{U} +  i_{K}\RU{}\right)  \\
                                &=& \frac{1}{\ep{0}}\Mp{U} \w \wt{U}\w i_{K}\RU{} - \frac{1}{\ep{0}\cc^{2}}\MM{U} \w i_{K}\J{} \w \wt{U} - \frac{1}{\ep{0}\cc}\MM{U} \w  i_{K}\RU{}  \\
                                &=& \left(\frac{1}{\ep{0}}\Mp{U} \w i_{K}\RU{} - \mu_{0}\MM{U} \w i_{K}\J{} \right)\w \wt{U} - \frac{1}{\ep{0}\cc}\MM{U}\w  i_{K}\RU{} \\
                                &=& \left(\frac{1}{\ep{0}}\Mp{U}(K)\RU{} + \mu_{0}i_{\wt{\Mm{U}}}\J{} \w \kappa \right)\w \wt{U} - \frac{1}{\ep{0}\cc}\MM{U}\w  i_{K}\RU{} \\
                                && \\
   F \w i_{K}j &=& \left( \Me{U} \w \wt{U} - \cc\MB{U} \right) \w \left(\frac{1}{\cc}i_{K}\J{} \w \wt{U} + i_{K}\RU{}\right) \\
               &=& \Me{U} \w \wt{U} \w i_{K}\RU{}  - \MB{U} \w i_{K}\J{} \w \wt{U} - \cc\MB{U} \w i_{K}\RU{} \\
               &=& \left(\Me{U} \w  i_{K}\RU{}  - \MB{U} \w i_{K}\J{}\right) \w \wt{U} - \cc\MB{U} \w i_{K}\RU{} \\
               &=& \left(\Me{U}(K)\RU{}  + i_{\wt{\Mb{U}}}\J{} \w \kappa\right) \w \wt{U} - \cc\MB{U} \w i_{K}\RU{} \\
               & & \\
   F \w \star i_{K}d\Pi &=& i_{K}d\Pi \w \star F = \left(\frac{1}{\cc}i_{K}\J{m} \w \wt{U} + i_{K}\RU{m}\right) \w \left( \ME{U} + \cc\Mb{U} \w \wt{U} \right) \\
                        &=& \frac{1}{\cc}i_{K}\J{m} \w \wt{U} \w \ME{U} + i_{K}\RU{m} \w \ME{U} + \cc i_{K}\RU{m} \w \Mb{U} \w \wt{U}   \\
                        &=& \left(\frac{1}{\cc}i_{K}\J{m} \w \ME{U} + \cc i_{K}\RU{m} \w \Mb{U}\right) \w \wt{U} + i_{K}\RU{m} \w \ME{U}  \\
                        &=& \left(-\frac{1}{\cc}i_{\wt{\Me{U}}}\J{m} \w \kappa + \cc \RU{m}\Mb{U}(K)\right) \w \wt{U} + i_{K}\RU{m} \w \ME{U}  \\
                        & & \\
   G \w i_{K}d\star\frac{\Pi}{\ep{0}} &=& -\frac{1}{\ep{0}}i_{K}G \w
   d\star\Pi =  -i_{K}F \w d\star\Pi - i_{K}\Pi \w d\star\frac{\Pi}{\ep{0}}\\
                        & & \\
   -i_{K}F \w d\star\Pi &=& -\left( i_{K}\Me{U} \w \wt{U} - \cc i_{K}\MB{U}\right) \w \left( -\frac{1}{\cc}\J{p} \w \wt{U} - \RU{p} \right) \\
                        &=& i_{K}\Me{U} \w \wt{U} \w \RU{p} -i_{K}\MB{U} \w \J{p} \w \wt{U} -  \cc i_{K}\MB{U} \w \RU{p} \\
                        &=& -\left(\Me{U}(K)\RU{p} + i_{K}\MB{U} \w \J{p} \right)\w \wt{U} - \cc i_{K}\MB{U} \w \RU{p} \\
                        &=& \left(-\Me{U}(K)\RU{p} - i_{\wt{\Mb{U}}}\J{p} \w \kappa \right)\w \wt{U} - \cc i_{K}\MB{U} \w \RU{p} \\
\end{eqnarray*}
\begin{eqnarray*}
   -i_{K}\Pi \w d\star\frac{\Pi}{\ep{0}} &=& -\frac{1}{\ep{0}}\left( i_{K}\Mp{U} \w \wt{U} - \frac{1}{\cc}i_{K}\MM{U} \right) \w \left( -\frac{1}{\cc}\J{p} \w \wt{U} - \RU{p}\right) \\
                                         &=& \frac{1}{\ep{0}}i_{K}\Mp{U} \w \wt{U} \w \RU{p} - \frac{1}{\ep{0}\cc^{2}}i_{K}\MM{U} \w\J{p} \w \wt{U} -  \frac{1}{\ep{0}\cc}i_{K}\MM{U} \w \RU{p} \\
                                         &=& -\left(\frac{1}{\ep{0}}\Mp{U}(K)\RU{p} + \mu_{0}i_{K}\MM{U} \w\J{p}\right) \w \wt{U} -  \frac{1}{\ep{0}\cc}i_{K}\MM{U} \w \RU{p} \\
                                         &=& \left(-\frac{1}{\ep{0}}\Mp{U}(K)\RU{p} - \mu_{0}i_{\wt{\Mm{U}}}\J{p} \w \kappa\right) \w \wt{U} -  \frac{1}{\ep{0}\cc}i_{K}\MM{U} \w \RU{p}.
\end{eqnarray*}
These expressions may be combined to produce the required 4-form
$d\tau_{K}^{SM}$ as described in \S 4, paper I.

%\newpage
%%%%%%%%%%%%%%%%%%%%%%%%%%%
%\begin{thebibliography}{99}

%%%%%%%%%%%%%%%%%%%%%%%
%\newpage

\end{document}